\documentclass[aip,jcp,reprint,superscriptaddress,longbibliography]{revtex4-1}
\usepackage{epsfig}
\usepackage{color}
\usepackage{amsmath}
\usepackage{amsfonts}
\usepackage{natbib}
\usepackage{notes2bib}

\newcommand{\erf}{\ensuremath{\mathop{\rm erf}}}
\newcommand{\erfc}{\ensuremath{\mathop{\rm erfc}}}
\newcommand{\rb}{\mathbf{r}}

\newcommand{\kb}{\mathbf{k}}
\newcommand{\avg}[1]{\left<#1\right>}
\newcommand{\len}[1]{\left|#1\right|}
\newcommand{\brac}[1]{\left[#1\right]}

\newcommand{\para}[1]{\left(#1\right)}

\newcommand{\rhoq}{\ensuremath{\rho^q}}

\newcommand{\V}{\ensuremath{\mathcal{V}}}
\newcommand{\Vr}{\ensuremath{\mathcal{V}_{\rm R}}}

\newcommand{\Vrl}{\ensuremath{\mathcal{V}_{\rm R1}}}

\newcommand{\kT}{\ensuremath{k_{\rm B}T}}

\newcommand{\Hb}{\ensuremath{\mathcal{H}}}

\newcommand{\Rbar}{\overline{\textbf{R}}}

\begin{document}



\title{Dielectric Response of Short and Long Range Models of Nonuniform Liquids and Implications for Machine Learned Interatomic Potentials}


\author{Atul C. Thakur}
\altaffiliation[Present Address: ]{NVIDIA Corp, 2788 San Tomas Expressway, Santa Clara, CA 95051}
\affiliation{Department of Chemistry and Chemical Biology, Rutgers University, Piscataway, NJ 08854}
\author{Harender S. Dhattarwal}
\altaffiliation[Present Address: ]{R. F. Smith School of Chemical and Biomolecular Engineering, Cornell University, Ithaca, NY, USA 14853}
\affiliation{Department of Chemistry and Chemical Biology, Rutgers University, Piscataway, NJ 08854}
\author{Richard C. Remsing}
\email[]{rick.remsing@rutgers.edu}
\affiliation{Department of Chemistry and Chemical Biology, Rutgers University, Piscataway, NJ 08854}



\begin{abstract}
Dielectric screening of electric fields is modified by interfaces and confinement, and any accurate description of interfacial chemistry necessitates properly modeling the nonuniform dielectric tensor. 
However, models with short range interactions, such as local machine learned interatomic potentials, are increasingly used in simulations of molecular interfaces without quantitative understanding of their nonuniform dielectric response. 
To build this understanding of nonuniform dielectric response, we use the conceptual framework provided by local molecular field (LMF) theory to quantify the roles of short and long range electrostatics in confined water. 
We find that short and long range models predict the same transverse (in-plane) dielectric profile but short range models predict longitudinal (perpendicular) dielectric profiles that are qualitatively different than their long range counterparts.
To understand these differences, we develop a Landau theory that shows that long range interactions stiffen polarization fluctuations by a factor of the bulk dielectric constant, reducing the interfacial polarization correlation length by an order of magnitude upon their inclusion. 
We also show that LMF theory, which captures long range interactions with a static averaged field, corrects the structure and dielectric response, but the standard fluctuation relations for the longitudinal dielectric constant no longer holds.
We derive a fluctuation relation and integral equation for LMF theory and demonstrate their accuracy for predicting longitudinal dielectric response. 
Machine learned interatomic potentials that incorporate interfacial configurations in their training essentially learn a static mean field correction at the interface, such that the integral equation can be used to predict the nonuniform dielectric profile from polarization fluctuations in short range, local machine learning models.
We also show that short range machine learned interatomic potentials can be combined with the symmetry preserving mean field (SPMF) theory to provide a linear-scaling alternative for handling long range interactions and predicting nonuniform dielectric functions.
\end{abstract}


\maketitle

\raggedbottom

\section{Introduction}
Screening of electric fields by polar molecules has profound consequences for chemistry in solution.
The extent of screening by polar molecules in an isotropic bulk fluid is quantified by the static dielectric constant, $\varepsilon$.
For water, $\varepsilon$ is a collective property that depends sensitively on molecular dipole moments, their correlations, and the description of molecular polarization, and its prediction has long been a demanding test of molecular models~\cite{MaddenChapter,Watanabe1989,SprikKlein1988,Zhang2023PRL}.
In a nonuniform system, however, the dielectric response is quantified by a spatially varying dielectric tensor. 
For a system with a slab-like symmetry, like solutions near membranes~\cite{Stern2003,Schneck2012PNAS,Kanduc2014}, mineral surfaces~\cite{Banuelos2023}, or metal electrodes~\cite{Limmer2013,Scalfi2021,Deissenbeck2021}, 
there are two non-zero components of the dielectric tensor~\cite{Stern2003,Ballenegger2005,Bonthuis2011PRL,Bonthuis2012Langmuir},
the transverse component parallel to the interface, $\varepsilon_\parallel(z)$,
and the longitudinal component perpendicular to the interface, $\varepsilon_\perp(z$.
The anisotropic interfacial dielectric tensor governs molecule-surface and surface-surface interactions through electrostatic forces, and an accurate description of the dielectric tensor is needed to interpret and predict interfacial chemistry, including double layer capacitance and interfacial solvation thermodynamics~\cite{Bonthuis2012Langmuir,Schlaich2016PRL,Fumagalli2018,Loche2020JPCB,Gonella2021}. 
The dielectric response of interfacial water is also reflected in the slow orientational and dielectric relaxation of water at the surfaces of self-assembled aggregates~\cite{Balasubramanian2002JPCB,Pal2004JCP} and in aqueous chemical dynamics confined within layered materials~\cite{Remsing2015JPCL,RemsingKlein2017CPL,Bhullar2021JMCA}.
The recent explosion of machine learned interaction potentials (MLIPs) has created a need to understand how general features of interatomic interactions influence the behavior of molecular systems, especially critical properties like the dielectric constant. 
Of particular interest are the relative roles of short and long range interactions.
Many MLIPs are inherently short range, because their development relies on an assumption of locality~\cite{Behler2007PRL,Bartok2010PRL,Zhang2018PRL,Anstine2023JPCA}. 
While the importance of long range interactions in empirical force fields has been appreciated for some time~\cite{Weeks2002ARPC,Widom1967Science,Weeks1971JCP,Chandler1983Science,ChenKaurWeeks2004,RodgersWeeks2008JPCM,RodgersWeeks2008PNAS,Remsing2016PNAS,Gao2023JPCB,Cox2020PNAS}, long range interactions in MLIPs are only beginning to be appreciated as methods for their description are developed~\cite{Grisafi2019JCP,Ko2021NatCommun,Grisafi2021ChemSci,Gao2022NatCommun,dhattarwal2023dielectric,Zhang2022DPLR,Cheng2025npj,King2025NatCommun,Loche2025JCP,Ramasubramanian2025arXiv, ko2025fast, balantrapu2026neutron}. 
For example, building on first principles descriptions of liquid water and aqueous solutions~\cite{Sun2016NatChem,Chen2017PNAS,Zhang2021JPCB,Zhang2022NatCommun}, neural network potentials that explicitly include long range electrostatics have been combined with neural network models for the centers of maximally localized Wannier functions to predict the dielectric permittivity of water~\cite{dhattarwal2023dielectric,Gao2022NatCommun,Zhang2022DPLR} and aqueous salt solutions~\cite{Zhang2023PRL}.
Much like their empirical predecessors, short range MLIPs lead to errors in energetics of the vapor phase and clusters~\cite{Yue2021JCP,Grisafi2019JCP,Cheng2025npj}, long range dipolar correlations in the bulk liquid~\cite{Cox2020PNAS,Gao2022NatCommun,Cheng2025npj}, polarization induced by electric fields~\cite{Gao2022NatCommun,Zhang2022DPLR}, and structure and screening at interfaces~\cite{Niblett2021JCP,Gao2022NatCommun,Cheng2025npj,Parker2026arXiv}.
At the heart of these properties is the collective response of the system to perturbations, and the constraints placed on this response by long range electrostatic interactions. 
As a result, it is imperative that we understand the role of long range electrostatics in the collective response of liquids in heterogeneous environments, exemplified by the nonuniform dielectric tensor. 
Here, we investigate the role of short and long range interactions in determining the components of the dielectric tensor of water confined between smooth surfaces as an ideal model of a nonuniform polar liquid. 
We work within the perspective provided by local molecular field (LMF) theory~\cite{ChenKaurWeeks2004,RodgersWeeks2008JPCM,Remsing2016PNAS,Gao2023JPCB}, in which short and long range interactions are separated over a physically chosen length scale, and the long range interactions are described by an effective static field. 
This allows us to clearly separate the effects of short range interactions from those determined by long range electrostatics, 
The one-dimensional (1D) field within LMF theory allows us to further quantify the importance of long range interactions parallel and perpendicular to the interface.
We also compare results from the static field in LMF theory with the fluctuating field provided by symmetry preserving mean field (SPMF) theory~\cite{Hu2014ChemCommun,Hu2022JCP} to quantify the role of fluctuations of the field in determining the dielectric response. 
We combine these simulation-based analyses with Landau theories for the short range, long range, and LMF systems to provide a model that illustrates the role of long range interactions in determining polarization and dielectric response in heterogeneous systems. 
We find that long range interactions are negligible in determining the transverse dielectric response parallel to the interface but critical to the longitudinal dielectric response perpendicular to the interface. 
Long range interactions can be efficiently described with SPMF and LMF theories, though LMF theory requires the use of a different fluctuation relation or integral equation to estimate the dielectric profile, which are derived here. 
The combination of these linear-scaling theories for long range electrostatics can be combined with local MLIPs to efficiently and accurately model nonuniform dielectric response by corrected short range models.
%

\begin{figure}[tb]
\begin{center}
\includegraphics[width=0.35\textwidth]{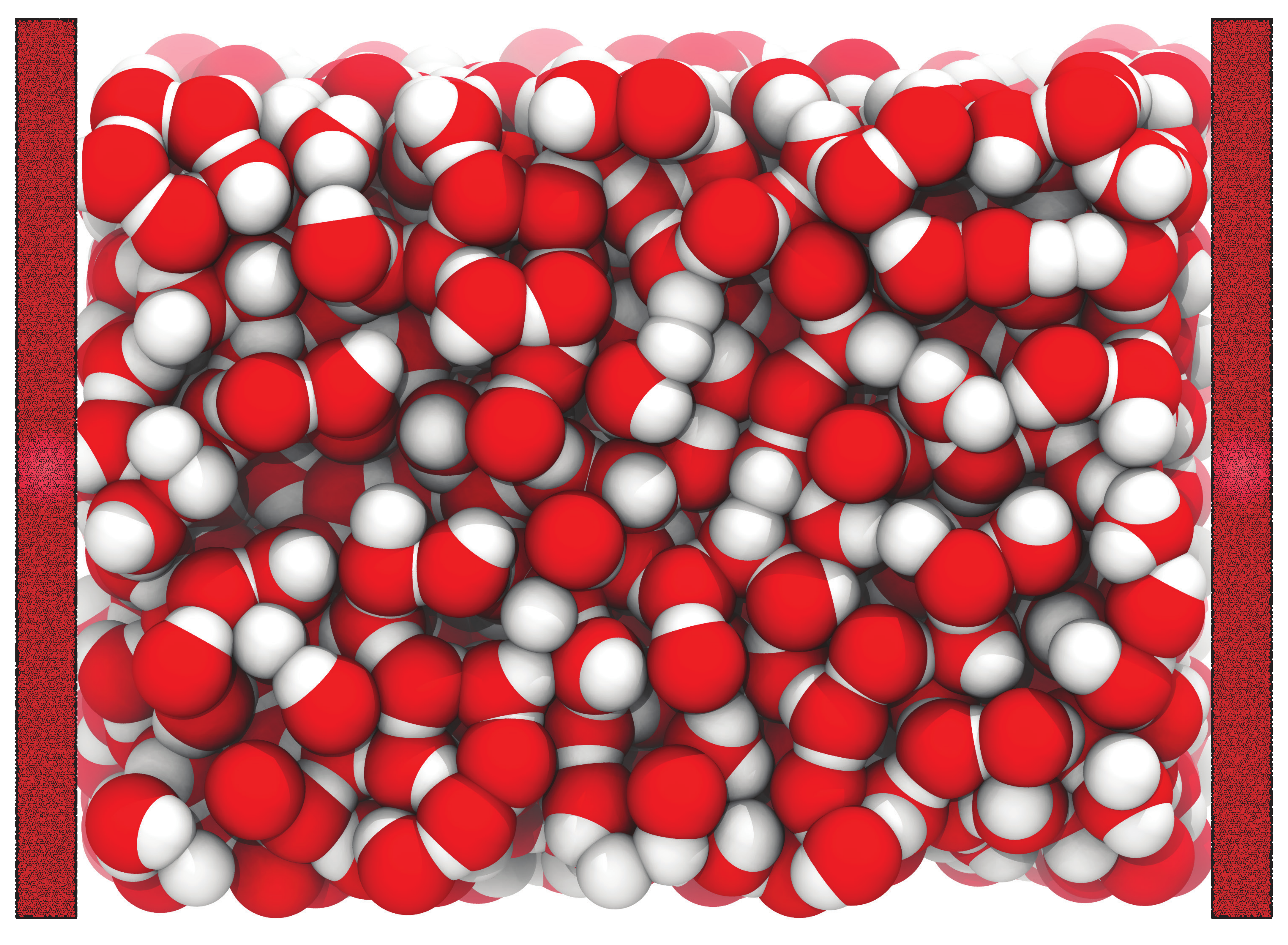}
\end{center}
\caption
{
Simulation snapshot showing the system with water confined between two hydrophobic Lee-Rossky walls~\cite{LeeRossky1994}. 
The oxygen atoms are colored red and the hydrogens are colored white. 
The $z$ axis is oriented along the direction of the confinement, perpendicular to the surface.
}
\label{fig:snap}
\end{figure}

\section{Models and Simulation Details}
To examine the effects of long range interactions on nonuniform dielectric response,
we must first define `short range' and `long range'.	
We consider Hamiltonians that can be written as a sum
of non-electrostatic terms (ne) and electrostatic (el) terms, 
$\Hb(\Rbar) = \Hb_{\rm ne}(\Rbar) + \Hb_{\rm el}(\Rbar)$.
Non-electrostatic interactions include intramolecular bonds and intermolecular
van der Waals interactions, for example. 
We focus on the electrostatic interactions, determined by
\begin{align}
\Hb_{\rm el}(\Rbar) &= \frac{1}{2}\sum_{i=1}^N \sum_{j=1}^N (1-\delta_{ij})\frac{q_i q_j }{4\pi\epsilon_0\len{\rb_i - \rb_j}} \\
			&\equiv   \frac{1}{2}\sum_{i=1}^N \sum_{j=1}^N (1-\delta_{ij})\frac{q_i q_j}{4\pi\epsilon_0} v(\len{\rb_i - \rb_j}) ,
\end{align}
where $q_i$ is the charge on atom $i$, $\delta_{ij}$ is the Kronecker delta function,
and $v(r)=1/r$.
We split the Coulomb potential into short range and long range parts
on the length-scale $\sigma$ according to
\begin{equation}
v(r) = v_0(r) + v_1(r) = \frac{\erfc(r/\sigma)}{r} + \frac{\erf(r/\sigma)}{r},
\end{equation}
respectively.
Using this separation, we exactly rewrite the Hamiltonian
as
\begin{equation}
\Hb(\Rbar) = \Hb_0(\Rbar) + \Hb_1(\Rbar), 
\end{equation}
where $\Hb_0(\Rbar)=\Hb_{\rm ne}(\Rbar)+\Hb_{\rm el,0}(\Rbar)$
includes all short range interactions and $\Hb_1(\Rbar)=\Hb_{\rm el,1}(\Rbar)$ is the contribution
from long range electrostatic interactions. 
Most treatments of electrostatics use this formal separation but differ
in their treatment of the long range $v_1(r)$ term. 
To understand the impact of different treatments of long range interactions on nonuniform dielectric response, we model water confined between smooth 9-3 Lennard-Jones walls~\cite{LeeRossky1994}, Fig.~\ref{fig:snap},
located at $\pm\frac{L}{2} = 21.53$~\AA.
Water is modeled using the flexible extended simple point charge (SPC/E) model~\cite{Berendsen1987,Wu2006SPCFw,Yue2021JCP}.
Simple point charge models of this type reproduce many of the structural, dielectric, and transport properties of liquid water~\cite{Jorgensen1983,Watanabe1989,Balasubramanian1996}.
All simulations were carried out using the LAMMPS simulation package\cite{Thompson2022LAMMPS}.
The system consisted of 1024 water molecules in a box with dimensions $L_x=27.72$~\AA, $L_y=27.72$~\AA, and $L_z=43.06$~\AA. 
The starting configuration was equilibrated at 300 K for at least 5 ns in the canonical (NVT) ensemble using three Nos{\'e}-Hoover chains~\cite{Martyna1992}. 
The production runs retained the same thermostat settings (three Nos{\'e}-Hoover chains, 100 fs damping constant), and used 1 fs integration timestep throughout.
All production simulations were run for longer than 80 ns unless otherwise mentioned. 

Our standard for accurate handling of long range electrostatic interactions is Ewald summation~\cite{deLeeuw1980}. 
Therefore, we refer to the system employing three-dimensional (3D) slab-corrected Ewald summation as the Full or long range (LR) system~\cite{YehBerkowitz1999}. 
The slab correction approximates Ewald summation with two-dimensional periodicity, for which specialized methods have been developed for planar surfaces and interfaces~\cite{HautmanKlein1992}.
Our model short range system neglects the long range Hamiltonian, $\Hb_1$, such that $v(r)$ is replaced by $v_0(r)$ in the evaluation of electrostatic interactions. 
Because $v_0(r)$ is the potential from a unit point charge screened by a unit Gaussian charge of opposite sign, we refer to this system as the Gaussian truncated (GT) model~\cite{RodgersWeeks2008JPCM,RodgersWeeks2008PNAS,Gao2023JPCB}. 
With a properly chosen $\sigma$ --- on the scale of typical nearest-neighbor distances ---
the GT model can accurately model the structure of bulk liquids; here we use $\sigma=4.5$~\AA. 
The accuracy of the GT model for bulk liquids is a consequence of force cancellation; the vector sum of the forces on a particle from long range electrostatics is zero to a good approximation in every relevant configuration~\cite{RodgersWeeks2008PNAS,Remsing2016PNAS}. 
As a result, the relevant forces determining bulk liquid structure are contained within the GT model,
and radial distribution functions for bulk water are nearly identical in LR and GT models, as well as for local MLIPs~\cite{Gao2023JPCB,RodgersHuWeeks2011,RodgersWeeks2008PNAS,Gao2022NatCommun,Cox2020PNAS}.
The force cancellation argument above also underlies the accuracy of local machine learned interaction potentials (MLIPs) for bulk fluids~\cite{Yue2021JCP,Gao2022NatCommun}. 
In these models, energies and/or forces are learned and computed based on the local coordination environment of an atom, typically defined by a spherical region on a length scale between $\sigma$ and $2\sigma$. 
Because long range forces tend to cancel, and the resulting long range energy is essentially a constant uniform background potential, a local MLIP trained on bulk liquid configurations is analogous to a short range GT model. 
To demonstrate this point, we developed a DeePMD model~\cite{Zhang2018PRL,Wang2018CPC,Zeng2023JCP} trained on configurations of flexible SPC/E water from our MD simulations of the LR system. 
The DP-Bulk model was trained using DeePMD-kit v2.1.1 with the smooth edition two-body embedding descriptor (\texttt{se\_e2\_a}) with a radial cutoff of 6~\AA. 
We used an embedding network of size (25, 50, 100) with axis neuron dimension 16 and a fitting network of size (240, 240, 240) with residual connections. 
Training used an exponentially decaying learning rate which decayed from $10^{-3}$ to $10^{-8}$ over $10^6$ optimizer steps, with an energy/force loss weighting that shifted emphasis from forces ($p_f = 1000 \rightarrow 10$) to energies ($p_e = 0.02 \rightarrow 1$) as training progressed.
No virial/stress contribution was included in the loss function ($p_v = 0$ throughout training). 
Both training and validation batch sizes were set to one and model validation was performed 100 steps over three validation batches. 
For DP-Bulk, the training dataset consisted of 3200 configurations (1600 at 300 K, 1600 at 2000 K) sampled from Ewald (LR) simulations of bulk flexible SPC/E water.
We will see below that any deficiencies of the GT model are also exhibited by this bulk DeePMD model (DP-Bulk). 
To illustrate this point in a more controlled manner, we also train a bulk model using GT interactions with $\sigma=4.5$~\AA, resulting in the DP-GT-Bulk model. 
The details of the training and dataset are the same as for the DP-Bulk model, except that the full interaction energies and configurations are replaced those generated by simulations of the GT model of flexible SPC/E water.
Finally, we also train a DeePMD model that incorporates knowledge of the interface. 
The model used an identical architecture and training protocol, but was instead trained on 4800 configurations (2400 at 300 K, 2400 at 2000 K) sampled from Ewald simulations of the confined system, allowing the model to additionally learn the local environments present near the interface.
The resulting DP-Int model should provide improved accuracy for the interfacial system of interest here. 
All other training hyperparameters were identical across the three models.
\section{Accurate longitudinal dielectric screening needs long range electrostatics}

\begin{figure*}[tb]
\begin{center}
\includegraphics[width=0.9\textwidth]{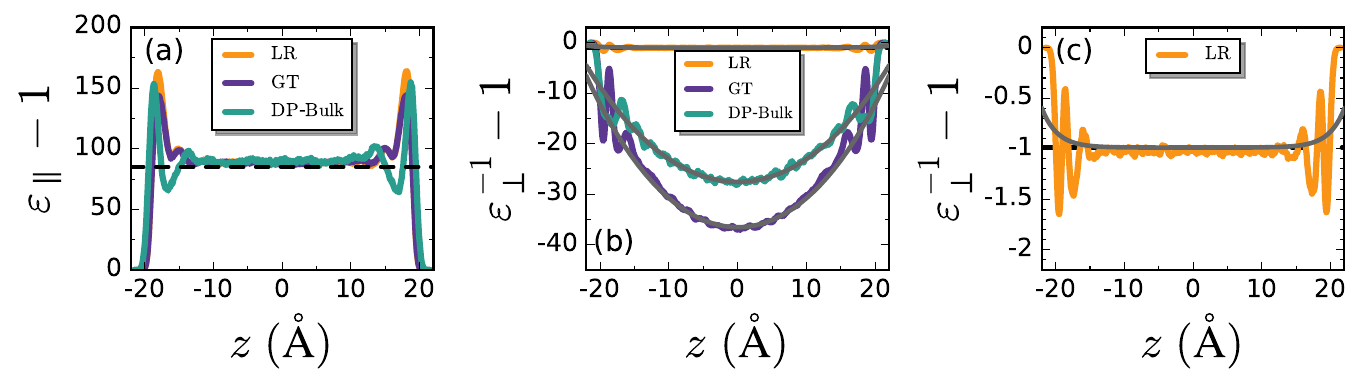}
\end{center}
\caption
{
(a) Transverse and (b) longitudinal dielectric constant predicted by the full long range (LR) model, the Gaussian truncated (GT) model, and the DeePMD model trained on bulk configurations (DP-Bulk). 
Gray lines are predictions from the mean field Landau theory. 
(c) Longitudinal dielectric constant and mean field predictions for the LR model on a zoomed in scale. 
}
\label{fig:GTeps}
\end{figure*}

%
In a nonuniform system, screening is described by a dielectric tensor, and in a planar geometry, this tensor involves two components: a transverse or parallel dielectric constant, $\varepsilon_{\parallel}(z)$, and a longitudinal or perpendicular dielectric constant, $\varepsilon_\perp(z)$.
We evaluate these nonuniform dielectric constants using the linear response framework of Netz and coworkers\cite{Bonthuis2011PRL,Bonthuis2012Langmuir,Schlaich2016PRL,Stern2003,Ballenegger2005},
\begin{equation}
\varepsilon_\parallel(z) = 1 + \beta \varepsilon_0^{-1} \avg{\delta m_\parallel(z,\Rbar) \delta M_\parallel(\Rbar)}_0
\label{eq:netzpar}
\end{equation}
and
\begin{equation}
\varepsilon_\perp^{-1}(z) = 1-\beta\varepsilon_0^{-1} \avg{\delta m_\perp(z,\Rbar)) \delta M_\perp(\Rbar)}_0,	
\label{eq:netzperp}
\end{equation}
where $\beta^{-1}=\kT$ is the inverse of Boltzmann's constant times the temperature,
$\varepsilon_0$ is vacuum permittivity, and $\avg{\cdots}_0$ indicates an ensemble
average in the absence of an applied external electrostatic field.
The transverse polarization is $m_\parallel(z,\Rbar))= \mp \rho_{\rm S}(z,\Rbar)$, where $\rho_{\rm S}(z,\Rbar)$
is the surface charge density obtained by making virtual cuts through the
configuration $\Rbar$ of the simulated system.
The longitudinal polarization is 
$m_\perp(z,\Rbar) = -\int_{-\infty}^z dz' \rhoq(z',\Rbar)$, where $\rhoq(z,\Rbar)$ is the instantaneous charge density. 
The total transverse ($\alpha=\parallel$) or longitudinal ($\alpha=\perp$) polarization is given by $M_\alpha(\Rbar)=A\int_{-\infty}^{\infty}m_\alpha(z,\Rbar) dz$, where $A$ is the surface area,
Within this formulation, predicting the dielectric constant requires an accurate
description of structural (polarization) fluctuations within the liquid. 
In the lateral direction along the surface (within the $xy$ plane), we expect a bulk-like force cancellation picture, such that long range forces cancel to a good approximation in every relevant configuration. 
As a result, structure, and the structural fluctuations determining $\varepsilon_\parallel(z)$, are determined mainly by short range interactions, and we might expect short range models to describe in-plane, transverse dielectric response, as they do in the bulk~\cite{BuiCox2025JPCM,Cox2020PNAS,BuiCox2026NatCommun}.
Indeed, the LR, GT, and DP-Bulk models all produce similar $\varepsilon_\parallel(z)$, Fig.~\ref{fig:GTeps}a.
The accuracy of the force cancellation argument decreases with decreasing dimensionality, such that some differences between the LR and GT/DP-Bulk models appear close to the walls, but otherwise, $\varepsilon_\parallel(z)$ is produced quantitatively by the short range models. 
The LR $\varepsilon_\perp^{-1}(z)$ oscillates near the interface but quickly plateaus to a value of $\varepsilon^{-1}$, where $\varepsilon$ is the static bulk dielectric constant, Fig.~\ref{fig:GTeps}b,c.
In contrast to the transverse direction, long range forces perpendicular to an interface do not cancel, and a net unbalanced force develops that has a significant impact on structure and fluctuations. 
Therefore, one might anticipate that the GT model will produce a significantly different $\varepsilon_\perp^{-1}(z)$ than the LR model. 
Indeed, $\varepsilon_\perp^{-1}(z)$ for the GT and DP-Bulk models become very negative, and both exhibit a nearly quadratic form with no plateau in the bulk, Fig.~\ref{fig:GTeps}b.
We can understand the different screening behavior of short and long range models through the development of a mean field Landau theory for polarization and dielectric response~\cite{Marcelja1976,Kanduc2014,Prelesnik2021PNAS,Limmer2024Book,Chandler1993PRE}.
To develop the theory, we write the long range part, $\Hb_1$, of our total effective Hamiltonian, $\Hb=\Hb_0+\Hb_1$, as
\begin{equation}
\Hb_1(\Rbar) = \frac{1}{8\pi\epsilon_0} \int d\rb \int d\rb' \rhoq(\rb;\Rbar) v_1(\len{\rb-\rb'}) \rhoq(\rb';\Rbar).
\end{equation}
In a slab geometry, we can split the charge density into transverse and longitudinal modes, allowing us to isolate the longitudinal part of the Hamiltonian, 
\begin{equation}
\Hb_1(\Rbar) =  \Hb_{1,\parallel}(\Rbar) + \Hb_{1,\perp}(\Rbar), 
\end{equation}
where the transverse component involves a summation over non-zero parallel wavevectors ($\kb_\parallel$), the longitudinal part is
\begin{equation}
\Hb_{1,\perp}(\Rbar) = \frac{A}{2} \int dz \int dz' \rhoq(z;\Rbar)G_1(\len{z-z'}) \rhoq(z';\Rbar), 
\end{equation}
and $G_1(z)$ is the appropriate one-dimensional interaction potential obtained by integrating $v_1(\rb)$ over the $xy$-plane. 
Noting that $\rhoq(z)=-\frac{\partial}{\partial z} m_\perp(z)$ and $\frac{\partial^2}{\partial z^2} G_1(z)=-\frac{1}{\epsilon_0} \rho_G(z)$, where $\rho_G(z)=\frac{1}{\sigma\sqrt{\pi}}e^{-z^2/\sigma^2}$ is a Gaussian density, we can integrate $\Hb_{1,\perp}$ by parts twice to write it in terms of the longitudinal polarization field,
\begin{equation}
\Hb_{1,\perp}(\Rbar) = \frac{A}{2\epsilon_0} \int dz \int dz' m_{\perp}(z;\Rbar) \rho_G(\len{z-z'}) m_{\perp}(z';\Rbar).
\end{equation}
The full system treats $\Hb_1$ in full detail, here with Ewald summation techniques, while the GT system neglects $\Hb_1$ completely, such that $\Hb^{\rm GT}=\Hb_0$. 
We focus on the longitudinal response only, because the transverse dielectric response is described well by all models studied. 
We then consider coarse-graining the system over some length scale on the order of $\sigma$ or larger, and approximate $\Hb_1$ by expanding $m(z')$ about $z$ to second order and integrating over $z'$,
\begin{align}
\Hb_{1}[m] &\approx  \frac{A}{2\epsilon_0} \int dz \brac{m(z)^2 - \frac{\sigma^2}{4} \para{\frac{\partial m}{\partial z}}^2}
\end{align}
where we have dropped the configurational notation for simplicity and used $m(z)$ to indicate the coarse-grained longitudinal polarization field.
In practice, this polarization field can be obtained through a Gaussian smoothing over the coarse-graining length, e.g. $\sigma$, such that
\begin{equation}
m(z) \approx m_\perp^\sigma(z) = \int dz' m_\perp(z')\rho_G(\len{z-z'}),
\end{equation}
which can be readily computed from molecular simulations~\cite{RemsingWeeks2015JPCB,Prelesnik2021PNAS}. 
We can write the short range effective Hamiltonian as~\cite{Marcelja1976,Kanduc2014,Prelesnik2021PNAS,Limmer2024Book,Chandler1993PRE},
\begin{equation}
\Hb_0[m] = A\int dz \brac{ \frac{1}{2\chi_0} m(z)^2 + \frac{\kappa_0}{2} \para{\frac{\partial m}{\partial z}}^2},
\end{equation}
where $\chi_0$ is the longitudinal susceptibility and $\kappa_0$ is related to the decay of correlations in the short range system. 
Finally, we consider the coupling of the system to a uniform constant external field. 
The full, long range system will couple to the displacement field $D$, while the short range systems couple to the bare field, $E_0$.
Therefore, the coupling of the system to the field can be written generally as
\begin{equation}
\Delta \Hb[m] = -f M_\perp = -f A\int dz m(z),
\end{equation}
where $f=D/\epsilon_0$ for the long range system and $f=E_0$ for the short range system. 
Because $\Hb_0$ and $\Hb_1$ are quadratic in $m$, we can write a general Hamiltonian
\begin{equation}
\Hb[m] = A \int dz \brac{\frac{1}{2\chi} m(z)^2 + \frac{\kappa}{2} \para{\frac{\partial m}{\partial z}}^2 - fm(z)}.
\end{equation}
In the long range system, $\chi^{-1} = \chi_0^{-1} + \epsilon_0^{-1}$ and $\kappa=\kappa_0 - \sigma^2/4\epsilon_0$.
In the short range system, $\chi^{-1}=\chi_0^{-1}$ and $\kappa=\kappa_0$.
We can compute a mean field theory for the polarization by minimizing the effective Hamiltonian, $\frac{\delta \Hb}{\delta m(z)} = 0$, which yields
\begin{equation}
\frac{1}{\chi} m(z) - \kappa \frac{\partial^2 m(z)}{\partial z^2} = f.
\end{equation}
The solution gives the mean polarization induced by the field $f$~\cite{Prelesnik2021PNAS},
\begin{equation}
\avg{m(z)}_f=\avg{m(z)}_0 + \chi f n(z),
\label{eq:mf}
\end{equation}
where 
\begin{align}
n(z;L,\lambda) &= 1-\frac{\cosh(z/\lambda)}{\cosh(L/2\lambda)},
\end{align}
and $\avg{\cdots}_f$ indicates an ensemble average in the presence of the field $f$.
The polarization in the absence of an applied field is induced by structuring at the walls and is given by~\cite{Prelesnik2021PNAS}
\begin{equation}
\avg{m(z)}_0 = -m_w \frac{\sinh(z/\lambda)}{\sinh(L/2\lambda)},
\label{eq:m0}
\end{equation}
where $\mp m_w = \avg{m(\pm L/2)}$ is the value of the polarization at the walls. 
The correlation length, $\lambda=\sqrt{\kappa\chi}$, describes the decay of the polarization from the walls into the bulk, and reduces to 
$\lambda_0=\sqrt{\kappa_0 \chi_0}$ in the short range system.
Using the relationship between the susceptibility and the dielectric constant, $\varepsilon-1=\chi_0/\epsilon_0$, the correlation length of the short range system can be related to that of the long range system, $\lambda_0^2=\varepsilon \lambda^2 + (\varepsilon-1)\sigma^2/4$.
However, $\lambda$ is a molecular length scale that is coarse-grained over in Landau theory, such that we can approximate the correlation length of the short range models as
\begin{equation}
\lambda_0 \approx \frac{\sigma}{2}\sqrt{\varepsilon-1}.
\end{equation}
To determine the polarization and dielectric profiles of the LR models from Landau theory, we estimate the correlation length for the coarse-grained polarization as $\lambda\approx \lambda_0 /\sqrt{\varepsilon}$.
%

\begin{figure}[tb]
\begin{center}
\includegraphics[width=0.4\textwidth]{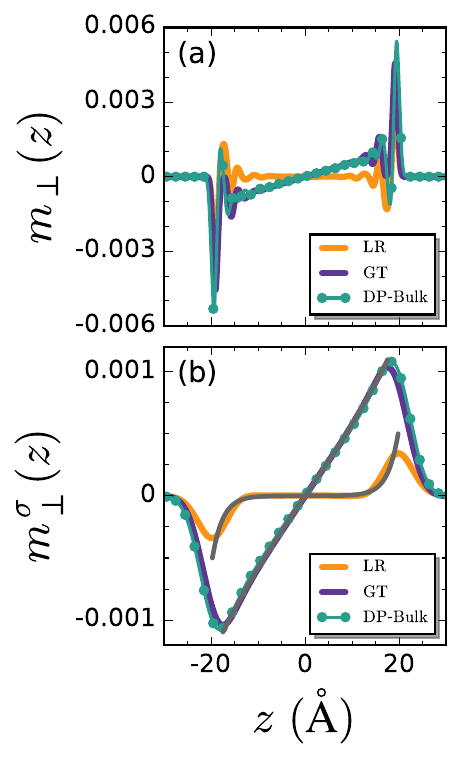}
\end{center}
\caption
{
(a) Bare and (b) Gaussian smoothed longitudinal polarization for the LR and GT models. Gray lines in (b) are fits from the mean field theory. 
}
\label{fig:pol}
\end{figure}

%
The theoretical form of $\avg{m(z)}_0$ is consistent with our simulation results. 
For a smoothing length of $\sigma=4.5$~\AA, the correlation lengths are $\lambda_0\approx20$~\AA \ and $\lambda\approx 2$~\AA \ with the determined static dielectric constant of $\varepsilon=86$ for the bulk water model. 
The only adjustable parameter is the position of the walls within the coarse-grained Landau theory, which we treat as a fit parameter, i.e. $L_{\rm eff}=L+2\delta$, and we find $\delta\approx\sigma/2$.
The Gaussian smoothed polarization, $\avg{m_\perp^\sigma(z)}$, agrees with Eq.~\ref{eq:m0} for both the LR and GT systems, until the walls are reached, despite the polarization being very different in each system. 
In the LR system, water polarizes near the walls, but this polarization is rapidly screened and zero polarization is achieved in the bulk. 
In contrast, the polarization that develops near the wall in the GT system is not screened on the lengthscale of the system and an approximately linear polarization profile develops across the system. 
Interfacial polarization is damped over much larger scales in the short range models. 
%

\begin{figure}[tb]
\begin{center}
\includegraphics[width=0.5\textwidth]{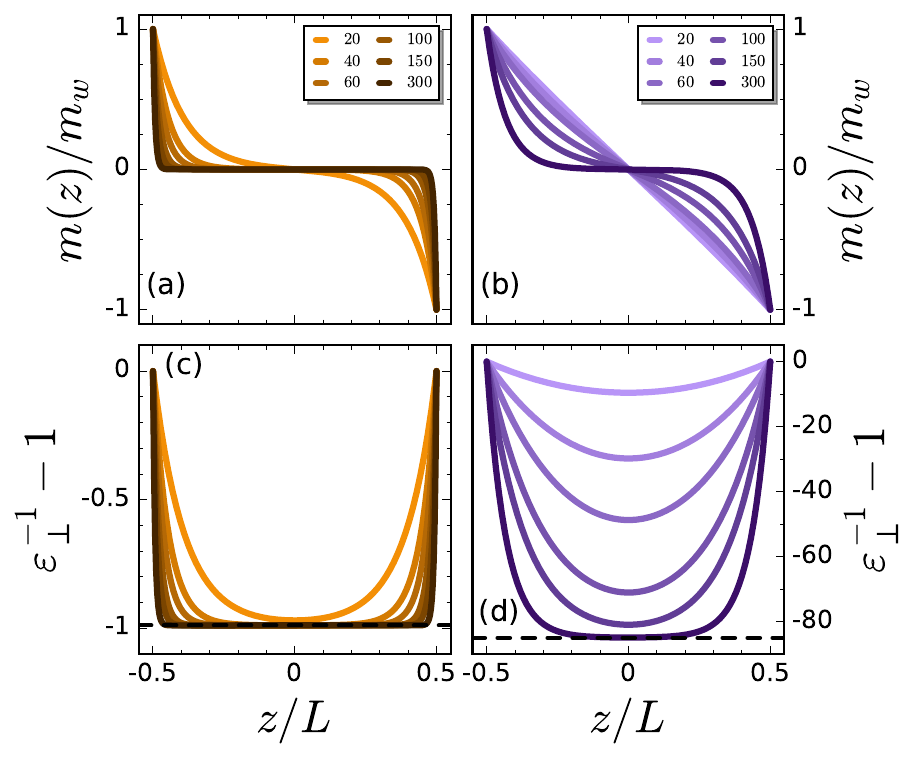}
\end{center}
\caption
{
(a,b) Predicted polarization and (c,d) longitudinal dielectric for (a,c) the LR model and (b,d) GT model as a function of the distance between confining walls, $L$. Black dashed line indicates the bulk dielectric constant.
}
\label{fig:size}
\end{figure}

\begin{figure}[b]
\begin{center}
\includegraphics[width=0.5\textwidth]{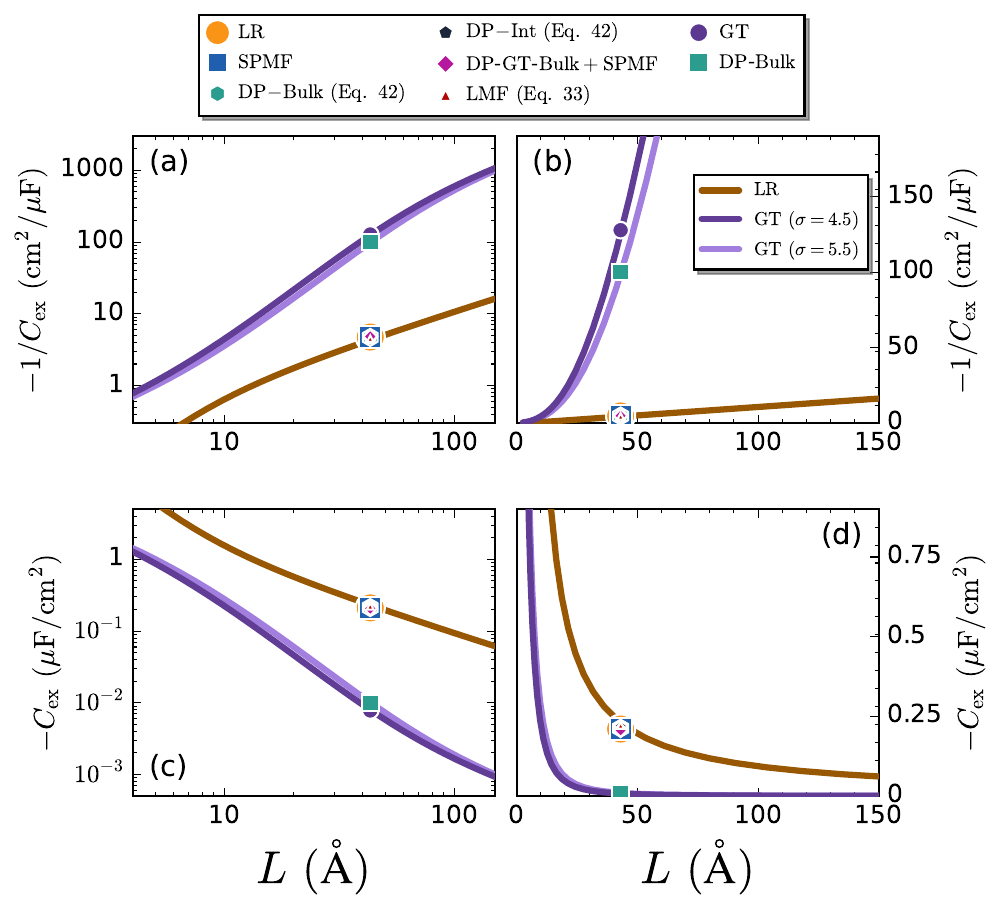}
\end{center}
\caption
{
(a,b) Inverse of the excess capacitance and (c,d) the excess capacitance as a function of $L$ predicted by the Landau theory for the long range model and for the short range model with $\sigma=4.5$~\AA \ for comparison with GT simulations and $\sigma=5.5$~\AA \ for comparison to the DP-Bulk MLIP results. 
Data points indicate numerical integrals of $\varepsilon_\perp^{-1}(z)$ for the various models, all at the same $L$. 
(a) and (b) are the same data on log and linear scales, as are (c) and (d), respectively. 
}
\label{fig:cex}
\end{figure}

%
We can estimate the change in polarization induced by the field from polarization fluctuations in the absence of the field using linear response theory,
\begin{equation}
\avg{\Delta m(z)}_f = \beta f \avg{\delta m(z;\Rbar) \delta M(\Rbar)}_0,
\label{eq:lrt}
\end{equation}
where $\avg{\Delta m(z)}_f \equiv  \avg{m(z;\Rbar)}_{f} - \avg{m(z;\Rbar)}_{0}$.
These fluctuations are also related to the longitudinal component of the dielectric constant through Eq.~\ref{eq:netzperp}, such that 
\begin{equation}
\varepsilon_\perp^{-1}(z) = 1-\frac{\chi}{\epsilon_0} n(z).
\end{equation}
The definitions of $\chi$ for the short and long range systems, combined with $\varepsilon-1=\chi_0/\epsilon_0$, yields expressions for the longitudinal dielectric constant profiles. 
For short range models, this produces
\begin{equation}
\varepsilon_\perp^{-1}(z) -1 = -(\varepsilon-1) n(z;L,\lambda_0),
\label{eq:sreps}
\end{equation}
while the inclusion of long range interactions leads to
\begin{equation}
\varepsilon_\perp^{-1}(z) -1 = -\para{\frac{\varepsilon-1}{\varepsilon}}n(z;L,\lambda).
\label{eq:lreps}
\end{equation}
The prefactor is larger by a factor of $\varepsilon$ in the short range system, consistent with previous work~\cite{Cox2020PNAS,BuiCox2025JPCM,BuiCox2026NatCommun}
Moreover, the decay of the polarization is significantly longer in the short range system, consistent with the polarization profiles and implying that $\varepsilon_\perp^{-1}(z)$ has a significant $L$-dependence in short range models.
The GT and LR dielectric profiles are described very well by Eq.~\ref{eq:sreps} and Eq.~\ref{eq:lreps}, Fig.~\ref{fig:GTeps}b,c.
The DP-Bulk model produces a $\varepsilon_\perp^{-1}(z)$ qualitatively similar to that of the GT model, suggesting that local MLIPs trained on bulk configurations contain the same physics as the GT model --- only short range interactions. 
Therefore, we can extract an effective $\sigma$ for the DP-Bulk model by fitting its dielectric profile to Eq.~\ref{eq:sreps}, Fig.~\ref{fig:GTeps}b.
The fit yields $\sigma_{\rm eff}\approx 5.5$~\AA, in excellent agreement with the 6~\AA \ radial cutoff used to define the descriptors of the DP-Bulk model, suggesting that local MLIPs behave in a manner consistent with GT electrostatics.
Finally, our theory enables the examination of the $L$-dependence of $\varepsilon_\perp^{-1}(z)$ and $m(z)$. 
Because $\lambda$ is small in the LR model, on the order of a molecular length scale, the resulting $\varepsilon_\perp^{-1}(z)$ and $m(z)$ rapidly reach a plateau in the bulk region, Fig.~\ref{fig:size}a,c.
In contrast, $\lambda_0$ is large in short range models, and the resulting $\varepsilon_\perp^{-1}(z)$ appears parabolic, Fig.~\ref{fig:size}d.
This is a result of $\lambda_0$ being comparable or larger than $L$ --- if the polarization from a single wall decays over a scale $\lambda_0$, there can not be a bulk region if $L$ is of order $2\lambda_0$ or less. 
In this limit, $n(z)$ can be expanded as $n(z)\sim \brac{L^2/4-z^2}/2\lambda_0^2$, which is the expected parabola.
However, when $L$ is significantly larger than $\lambda_0$, $\varepsilon_\perp^{-1}(z)-1$ indeed reaches a plateau in the bulk region at the expected bulk value of $1-\varepsilon$, Fig.~\ref{fig:size}d.
Similarly, the linear polarization profile that develops across the system is a result of overlapping interfacial regions with large correlation lengths, and large $L$ results in polarization profiles that reach zero in the bulk, Fig.~\ref{fig:size}b.
The drastic difference in the form of the dielectric profile can significantly impact estimates of the capacitance. 
To illustrate this point, we reference the capacitance, $C$, to its vacuum value, $C_{\rm vac}$, and focus on the excess capacitance,
\begin{equation}
\frac{1}{C_{\rm ex}} \equiv \frac{1}{C} - \frac{1}{C_{\rm vac}} = \frac{1}{\epsilon_0} \int dz \brac{\varepsilon_\perp^{-1}(z)-1}.
\label{eq:cex}
\end{equation}
For two plates separated by a distance $L$, the vacuum capacitance is $C_{\rm vac}^{-1} = L/\epsilon_0$. 
We can analytically integrate the dielectric profiles from Landau theory to get
\begin{equation}
\frac{1}{C_{\rm ex}^{\rm LR}} = -\frac{1}{\epsilon_0}\para{\frac{\varepsilon-1}{\varepsilon}}\brac{L-2\lambda \tanh\para{\frac{L}{2\lambda}}}
\end{equation}
and
\begin{equation}
\frac{1}{C_{\rm ex}^{\rm GT}} = -\frac{1}{\epsilon_0}\para{{\varepsilon-1}}\brac{L-2\lambda_0 \tanh\para{\frac{L}{2\lambda_0}}}
\end{equation}
for the long range and short range systems, respectively. 
We also numerically integrate the dielectric profiles obtained from simulation to compare with the theoretical predictions. 
The estimates of $C_{\rm ex}$ from simulation agree remarkably well with the theoretical predictions; even the difference between GT ($\sigma=4.5$~\AA) and the DP-Bulk ($\sigma_{\rm eff}=5.5$~\AA) is captured.
The magnitude of $C_{\rm ex}$ is significantly larger in the LR models, and the difference between the LR and GT results grows with increasing $L$. 

\section{Correcting short range models with static effective fields}

If short range models are desired for their efficiency and/or physical appeal, one can try to correct their interfacial behavior by developing approximations to the long range Hamiltonian, $\Hb_1$.
One can approximate the long range Hamiltonian with a static  averaged field, as in local molecular field (LMF) theory. 
The self-consistent electrostatic LMF potential in the presence of an external field, $E_{\rm ext}(z)$, is
\begin{equation}
\Vr(z) = \V(z) + \int dz' \avg{\rhoq(z')}_{\rm R} G_1(\len{z-z'}),
\end{equation}
where $\V(z)$ is the potential that gives rise to the external field, which can be split into short and long range parts, $\V(z)=\V_0(z)+\V_1(z)$. 
This field is determined such that the singlet (one-body) charge density, $\rhoq(z)$, in the LMF (or mimic) system is equal to that in the full LR system. 
As a result, the LMF potential is a description of the averaged unbalanced long range electrostatic forces that particles feel at an interface.
Because $\V_0(z)$ can potentially contain short range interactions in the external field, we can formally move that to the short range Hamiltonian $\Hb_0$ and focus on the long range portion of the LMF potential,
\begin{equation}
\Vrl(z) = \V_1(z) + \int dz' \avg{\rhoq(z')}_{\rm R} G_1(\len{z-z'}),
\end{equation}
such that LMF theory approximates the long range part of the Hamiltonian by
\begin{equation}
\Hb_1(\Rbar)\approx \Hb_1^{\rm LMF}(\Rbar) = A \int dz \rhoq(z,\Rbar) \Vrl(z).
\end{equation}
The field from the LMF potential, $E_{\rm R1}(z)$, can be obtained by taking the negative gradient of $\Vrl(z)$, which, upon integrating by parts, can be written in terms of the polarization,
\begin{equation}
E_{\rm R1}(z) = E_{\rm ext}(z) - \frac{1}{\epsilon_0} \avg{m_{\perp}^\sigma(z)}_{\rm R},
\end{equation}
where $\avg{\cdots}_{\rm R}$ indicates an ensemble average in the LMF mimic system.
When the external field is a constant displacement field, 
\begin{equation}
E_{\rm R1}(z) = \frac{D}{\epsilon_0} - \frac{1}{\epsilon_0} \avg{m_{\perp}^\sigma(z)}_{\rm R},
\end{equation}
making it clear that the self-consistent LMF portion of the field is the negative of the smoothed polarization. 
The LMF potential incorporates the self-consistent response of the fluid to polarization that develops at the interface, in essence incorporating information about the nonuniform system into the Hamiltonian to produce the correct one-body structure. 
%
 
\begin{figure}[tb]
\begin{center}
\includegraphics[width=0.4\textwidth]{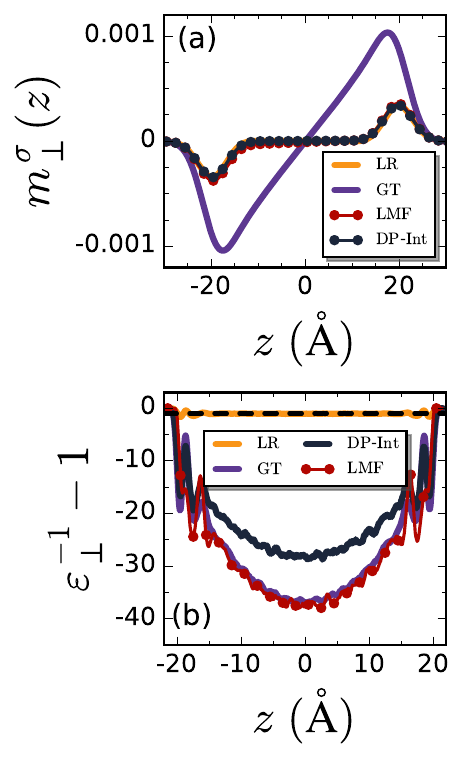}
\end{center}
\caption
{
(a) Gaussian smoothed polarization predicted by the LR, GT, LMF, and DP-Int models, and
(b) the corresponding longitudinal dielectric constants predicted by Eq.~\ref{eq:netzperp}. 
Dashed black line indicates the bulk dielectric constant. 
}
\label{fig:lmfpol}
\end{figure}

\begin{figure*}[tb]
\begin{center}
\includegraphics[width=0.99\textwidth]{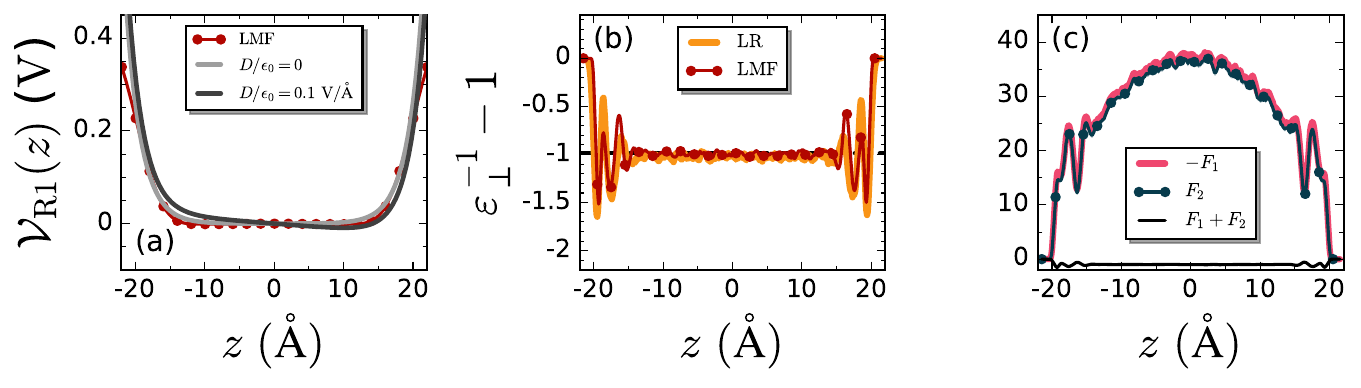}
\end{center}
\caption
{
(a) LMF potential from simulation at zero field and that predicted by the mean field theory at zero field and $D/\epsilon_0=0.1$~V/\AA. 
(b) Longitudinal dielectric constant for the LR model and from LMF theory via Eq.~\ref{eq:lmfeps}. 
(c) Fluctuation terms $F_1=-\beta \frac{1}{\epsilon_0} \avg{\delta m(z;\Rbar) \delta M(\Rbar)}_{\rm R0}$ and $F_2=\frac{\beta}{D\epsilon_0} \avg{\delta m(z;\Rbar)\brac{\delta \Phi_{\rm S}(\Rbar)-\delta \Phi_{\rm S0}(\Rbar)}}_{\rm R0}$ and their sum, $F_1+F_2 = \varepsilon_\perp^{-1}(z)-1$. 
}
\label{fig:lmfeps}
\end{figure*}

\begin{figure}[tb]
\begin{center}
\includegraphics[width=0.45\textwidth]{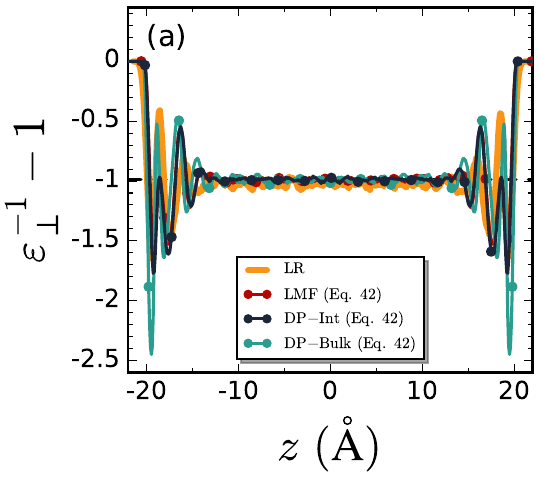}
\end{center}
\caption
{
Longitudinal dielectric profiles predicted using Eq.~\ref{eq:lmfinteq} for the LMF trajectory with $\sigma=4.5$~\AA \ and the DP-Int and DP-Bulk model trajectories with $\sigma=5.5$~\AA, as well as the results from the LR model using the standard fluctuation relation. 
}
\label{fig:inteq}
\end{figure}

%
Incorporating interfacial configurations into the training of a local MLIP is similar in spirit to this LMF framework. 
Indeed, MLIPs trained in this way are essentially learning the average LMF potential at the interface to correct interfacial structure~\cite{Niblett2021JCP}. 
Therefore, in the same way that we consider LMF theory for correcting the results of the short range GT model, we consider additional training of the DP-Bulk model that incorporates interfacial configurations to be a LMF-like correction of a short range model. 
We refer to the resulting MLIP trained on both bulk and interfacial configurations as the DP-Int model. 
The LMF and DP-Int polarization profiles are in excellent agreement with that of the LR model, demonstrating that both approaches correct the average structure at the interface, Fig.~\ref{fig:lmfpol}a.
The average field learned in both models provides a net torque on interfacial water molecules that corrects their orientation in a manner consistent with screening of the interfacial polarization layer. 
Despite this quantitative description of interfacial structure, applying Eq.~\ref{eq:netzperp} to compute $\varepsilon_\perp^{-1}(z)$, Fig.~\ref{fig:lmfpol}b, produces a dielectric constant profile similar to short range models!
The disagreement between the LR and LMF dielectric profiles suggests that LMF theory is unable to capture the necessary (two-body) fluctuations, which is a reasonable expectation for a one-body theory with an average field. 
This is consistent with the analysis of Cox~\cite{Cox2020PNAS}, who found that LMF theory captures the average polarization response of a water slab to an applied field but not the polarization fluctuations of the long range system.
However, the fluctuation-dissipation relation in Eq.~\ref{eq:netzperp} is derived for a system where the entire response to an applied field is contained in the fluid's polarization. 
In contrast, within LMF theory, the response to an external field is split between the induced polarization, $m(z)$, and the self-consistent LMF potential, $\Vrl(z)$.
Therefore, the fluctuation relation for $\varepsilon_\perp^{-1}(z)$ in Eq.~\ref{eq:netzperp} is not appropriate for LMF theory, and the relevant fluctuation relation must account for the response of the LMF potential to the external field. 
The relevant fluctuation relation is
\begin{align}
\varepsilon_\perp^{-1}(z) &= 1 - \frac{\beta}{\epsilon_0}  \avg{\delta m(z;\Rbar) \delta M(\Rbar)}_{\rm R0}  \nonumber \\
&+\frac{\beta}{D \epsilon_0} \avg{\delta m(z;\Rbar)\brac{\delta \Phi_{\rm S}(\Rbar)-\delta \Phi_{\rm S0}(\Rbar)}}_{\rm R0},
\label{eq:lmfeps}
\end{align}
where $\Phi_{\rm S}(\Rbar) = A \int dz \avg{m^\sigma(z;\Rbar)}_{\rm R} m(z)$ and $\Phi_{\rm S0}(\Rbar) = A \int dz \avg{m^\sigma(z;\Rbar)}_{\rm R0} m(z)$, and we derive this expression in Appendix~A.
The first fluctuation term is equivalent to the standard formula --- the nonuniform dielectric profile is proportional to the correlation between a fluctuation in the instantaneous polarization at $z$ and the total polarization.
Because LMF theory uses a self-consistent renormalized (or restructuring) field to produce the correct response to an external field, there is an additional term in $\varepsilon_\perp^{-1}(z)$ that involves the coupling of the instantaneous polarization at $z$ to the change in the LMF field due to the external applied field.
The field is specific to the particular value of $D$, such that the fluctuation term involving the LMF potentials scales with $D$, and a factor of $D^{-1}$ appears in its prefactor to cancel the field dependence. 
The explicit external field dependence of the LMF potential can make the fluctuation formula impractical --- simulations both in the presence and absence of the external field are required at least to compute the LMF potentials, and the dielectric constant can be readily evaluated from singlet densities in the presence of the field. 
We can demonstrate that LMF theory produces the correct response to the external field with our mean field Landau theory. 
The effective Hamiltonian, $\Hb[m] = \Hb_0[m] + \Hb_{\rm R}[m]$, uses the same short range form as before, but the long range interactions and the field term are contained in
\begin{align}
\Hb_{\rm R}[m] &= - A\int dz E_{\rm R1}(z) m(z) \\
= &-\frac{D}{\epsilon_0}A \int dz m(z) \nonumber \\
&+ \frac{A}{\epsilon_0} \int dz \bigg[\avg{m(z)}_{\rm R} m(z) \nonumber \\
&- \frac{\sigma^2}{4} \para{\frac{\partial m}{\partial z}} \para{\frac{\partial \avg{m(z)}_{\rm R}}{\partial z}}\bigg]
\end{align}
where we used $\avg{m^\sigma(z)}_{\rm R}=\int dz' \rho_G(\len{z-z'})\avg{m(z')}_{\rm R}$ and expanded $m(z')$ to second order about $z$.
Then, the total Hamiltonian to be minimized is
\begin{align}
\Hb[m]=A\int dz  \bigg[&\frac{m(z)^2}{2\chi_0} + \frac{\kappa_0}{2} \para{\frac{\partial m}{\partial z}}^2  \\
&- \para{ \frac{D}{\epsilon_0} - \frac{1}{\epsilon_0}\avg{m(z)}_{\rm R} }m(z) \nonumber \\
&- \frac{\sigma^2}{4\epsilon_0} \para{\frac{\partial m}{\partial z}} \para{\frac{\partial \avg{m(z)}_{\rm R}}{\partial z}} \bigg].
\end{align}
Minimizing the effective Hamiltonian, and noting that $m(z)=\avg{m(z)}_{\rm R}$ at the minimum, yields
\begin{equation}
\frac{1}{\chi} m(z) - \kappa \frac{\partial^2 m(z)}{\partial z^2} = \frac{D}{\epsilon_0},
\end{equation}
the same form as the long range model, which has the solution
\begin{equation}
m(z) = m_0(z) + \frac{D}{\epsilon_0} \chi n(z),
\end{equation}
where $\chi = (1/\chi_0 + 1/\epsilon_0)^{-1}$. 
Using our linear response relation, we can write the polarization change as
\begin{align}
\frac{D}{\epsilon_0} \chi n(z) &= \beta \frac{D}{\epsilon_0} \avg{\delta m(z;\Rbar) \delta M(\Rbar)}_{\rm R0} \nonumber \\
&- \frac{\beta}{\epsilon_0} \avg{\delta m(z;\Rbar)\brac{\delta \Phi_{\rm S}(\Rbar)-\delta \Phi_{\rm S0}(\Rbar)}}_{\rm R0},
\end{align}
such that using Eq.~\ref{eq:lmfeps} gives
\begin{equation}
\varepsilon_\perp^{-1}(z) = 1-\chi \frac{1}{\epsilon_0}n(z) = 1-\para{\frac{\varepsilon-1}{\varepsilon}}n(z),
\end{equation}
and the dielectric profile is the same as that in the full system. 
We can further show that LMF theory can produce the correct longitudinal dielectric constant profile explicitly by evaluating Eq.~\ref{eq:lmfeps}. 
To efficiently evaluate the fluctuation term involving the LMF potentials at zero and finite field, we estimate $\Vrl(z)$ from mean field theory. 
We fit the computed LMF potential at zero field to the theory and use the resulting parameters to estimate the field at $D/\epsilon_0=0.1$~V/\AA, Fig.~\ref{fig:lmfeps}a; the specific value of the field is unimportant as long as it is in the linear regime. 
We then use the theoretical LMF potentials to determine $\Phi_{\rm S}$ and $\Phi_{\rm S0}$.
The resulting $\varepsilon_\perp^{-1}(z)$ produced by Eq.~\ref{eq:lmfeps} is in excellent agreement with the LR results, Fig.~\ref{fig:lmfeps}b.
The correlation between polarization and LMF potential fluctuations is of the same order of magnitude of that between the polarization and the total polarization but opposite in sign, Fig.~\ref{fig:lmfeps}c.
As a result, these two terms largely cancel to produce the correct dielectric profile. 
In the context of interface-trained MLIPs, an analogue to Eq.~\ref{eq:lmfeps} must be used to estimate the longitudinal dielectric constant.  
To do so, a separate MLIP would be trained for each value of the field, such that each MLIP produces the correct interfacial structure by conceptually learning a field-specific LMF potential~\cite{Niblett2021JCP}. 
As a result, the relevant fluctuation formula for the dielectric constant would involve fluctuations in the total energy difference of the two MLIPs evaluated on the same configurations and practically involve resampling (postprocessing) the trajectory in the absence of the field with the MLIP trained with the field. 
This is impractical and directly computing the dielectric response from a single simulation in the presence of the field would be more efficient. 
The impracticalities of Eq.~\ref{eq:lmfeps} associated with the need for information obtained in the presence of the applied field can be overcome by rewriting the fluctuation relation as an integral equation,
\begin{align}
\varepsilon_\perp^{-1}(z) &= 1 - \frac{\beta}{\epsilon_0} \avg{\delta m(z;\Rbar) \delta M(\Rbar)}_{\rm R0}  \nonumber \\
&+\frac{\beta A}{\epsilon_0} \int dz' \int dz'' \avg{\delta m(z;\Rbar) \delta m(z';\Rbar)}_{\rm R0} \nonumber \\ &\times \rho_G(\len{z'-z''};\sigma)\para{1-\varepsilon_\perp^{-1}(z'')},
\label{eq:lmfinteq}
\end{align}
which we derive in Appendix~B and have highlighted the dependence of the Gaussian smoothing function on $\sigma$.
The integral equation is an Ornstein-Zernike (OZ) equation with a random phase approximation (RPA) closure that relates the short range reference system to the long range full system~\cite{HansenMcDonald2013,Andersen1972JCP}, as well as a classical analog of the Dyson equation and RPA integral equation for the inverse dielectric function of an electron gas~\cite{pines2018theory,Martin1959PR,Newns1970PRB,Horing1985PRB}, except here the reference system is not non-interacting but a fully interacting short range liquid. 
We have written Eq.~\ref{eq:lmfinteq} in a way that isolates the standard fluctuation relation, and the last term is equivalent to the LMF-based term in Eq.~\ref{eq:lmfeps}, but now written so that no field-dependent terms appear. 
The only input from the simulation is polarization fluctuations in the absence of the field. 
With this information, we can solve Eq.~\ref{eq:lmfinteq} for $\varepsilon_\perp^{-1}(z)$ through standard matrix inversion techniques. 
The resulting dielectric profile for the LMF system is in quantitative agreement with both the LR results and the results of the fluctuation relation Eq.~\ref{eq:lmfeps} (Fig.~\ref{fig:inteq}), demonstrating that the integral equation provides an accurate estimate of the dielectric profile from a single LMF-based simulation. 
The RPA closure is exact at long wavelengths, where the short and long range systems have the same bulk dielectric constant~\cite{Cox2020PNAS}.
This fixes the bulk plateau region of $\varepsilon_\perp^{-1}(z)$ at $1/\varepsilon$, so that any errors due to the approximation appear at the interface, as evidenced by the difference in the DP-Int and DP-Bulk data in Fig.~\ref{fig:inteq}.
Because of the constraint on the plateau in the bulk region, this part of the prediction is insensitive to $\sigma$, but the interfacial structure is sensitive to the particularly value of the smoothing length used in the calculation.
%

%
If interface-trained MLIPs like the DP-Int model are indeed learning something like a LMF potential to correct the predicted forces at the interface, then we should be able to use Eq.~\ref{eq:lmfinteq} to accurately estimate the longitudinal dielectric profile. 
The only additional information needed is the value of $\sigma$ to use in the Gaussian. 
We have an estimate of this smoothing length from fitting the standard fluctuation relation results to our Landau theory, $\sigma_{\rm eff}=5.5$~\AA. 
Solving Eq.~\ref{eq:lmfinteq} with the DP-Int polarization fluctuations produces a $\varepsilon_\perp^{-1}(z)$ in excellent agreement with the full LR model results, Fig.~\ref{fig:inteq}, and a correspondingly accurate capacitance, Fig.~\ref{fig:cex}. 
Because Eq.~\ref{eq:lmfinteq} was derived within the context of LMF theory, the good agreement of its results for DP-Int with the full model provides further support that interface-trained MLIPs learn effective potentials that try to account for the averaged effects of long range interactions at interfaces, similar to LMF theory.
Perhaps more importantly, when combined with an estimate of $\sigma$ from fitting uncorrected dielectric profiles, Eq.~\ref{eq:lmfinteq} enables accurate estimates of the long range longitudinal dielectric profiles from short range models trained on interfacial configurations.
We can further push the applicability of Eq.~\ref{eq:lmfinteq} to a MLIP trained only on bulk configurations, which corresponds to approximating the polarization correlations by the zeroth order term in a response theory, $\avg{\delta m(z;\Rbar) \delta m(z';\Rbar)}_{\rm R0}\approx \avg{\delta m(z;\Rbar) \delta m(z';\Rbar)}_{\rm 0}$.
Despite the simplicity of this approximation, the resulting dielectric profile is in excellent agreement with that of the LR model everywhere except immediately near the walls, Fig.~\ref{fig:inteq}. 
This can be expected, because near the wall is precisely where the effective field is strongest, Fig.~\ref{fig:lmfeps}a, and any influence of that field is neglected in this zeroth order approximation. 
We anticipate that the deviations of this corrected DP-Bulk dielectric profile near the walls can be corrected by including the first order term in the approximation of the polarization correlations to produce a linear response theory, which would involve fluctuations in the interactions with a LMF-like potential. 
However, even at this level of a zeroth order approximation to Eq.~\ref{eq:lmfinteq}, the dielectric profile of the DP-Bulk model produces a capacitance in agreement with the full model, Fig.~\ref{fig:cex}, and Eq.~\ref{eq:lmfinteq} can be used to estimate quantitative dielectric profiles from short range models.

\section{Correcting short range models with fluctuating effective fields}

\begin{figure}[tb]
\begin{center}
\includegraphics[width=0.4\textwidth]{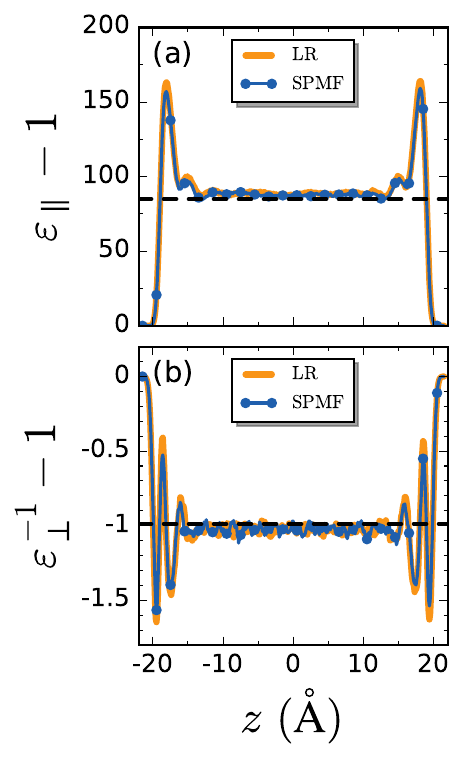}
\end{center}
\caption
{
(a) Transverse and (b) longitudinal dielectric constants as predicted by the LR model and SPMF theory. Black dashed line indicates the bulk dielectric constant. 
}
\label{fig:spmf}
\end{figure}

%
If we want to avoid corrections from static fields and/or training on interfaces altogether, while also increasing the efficiency of evaluating long range interactions in MLIPs in slab geometries, we can combine a local MLIP with the linear-scaling symmetry preserving mean field (SPMF) theory~\cite{Hu2014ChemCommun,Yi2017PCCP,Hu2022JCP}.
SPMF theory essentially neglects the transverse component of the long range interactions --- leveraging the force cancellation concept --- to approximate the long range Hamiltonian as
\begin{equation}
\Hb_1(\Rbar) \approx \Hb_1^{\rm SPMF}(\Rbar) = \Hb_{1,\perp}(\Rbar)).
\end{equation}
By treating the 1D long range electrostatics in each configuration, SPMF theory accurately models longitudinal electrostatic fluctuations. 
As a result, we can expect SPMF theory to produce accurate estimates of both the longitudinal and transverse dielectric constant. 
Indeed, if we apply SPMF to the GT model, the resulting $\varepsilon_\parallel(z)$ and $\varepsilon_\perp^{-1}(z)$ are in quantitative agreement with those of the LR system, Fig.~\ref{fig:spmf}.
%

\begin{figure}[tb]
\begin{center}
\includegraphics[width=0.5\textwidth]{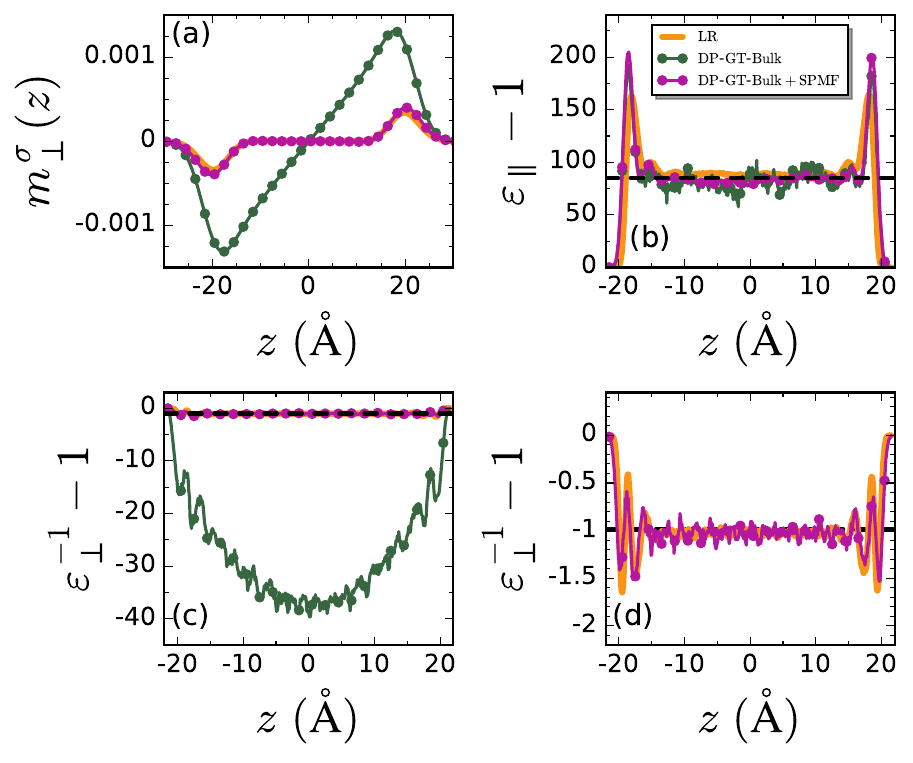}
\end{center}
\caption
{
(a) Smoothed polarization, (b) transverse dielectric constant, and (c,d) longitudinal dielectric constant for the LR, DP-GT-Bulk, and DP-GT-Bulk model with SPMF theory. 
Panel (d) is the same as (c) but on a different scale. 
}
\label{fig:dpgtbulk}
\end{figure}

%
To apply SPMF to a MLIP in a controllable manner, we train a DeePMD model on bulk configurations of the GT model of flexible SPC/E water with $\sigma=4.5$~\AA. 
We refer to the model as DP-GT-Bulk. 
In the interfacial system, the DP-GT-Bulk produces structure and dielectric constants consistent with the GT model and qualitatively consistent with the DP-Bulk model, Fig.~\ref{fig:dpgtbulk}a-c. 
To correct the behavior of the DP-GT-Bulk model, we can evaluate long range interactions within SPMF theory, using the point charges of the flexible SPC/E model.
For MLIPs trained on first principles data, where fixed point charges are not available, the charge density needed to evaluate the SPMF field could instead be obtained from a model for the positions of Wannier centers, as in neural network potentials with explicit long range electrostatics~\cite{Gao2022NatCommun,dhattarwal2023dielectric,Zhang2022DPLR,Zhang2023PRL}.
The resulting combination of DP-GT-Bulk and SPMF theory indeed leads to the correct polarization profiles, $\varepsilon_\parallel(z)$, and $\varepsilon_\perp^{-1}(z)$, Fig.~\ref{fig:dpgtbulk}a-d. 
Finally, we note that accurate predictions of the capacitance result from the corrected $\varepsilon_\perp^{-1}(z)$. 
The dielectric profile produced by LMF theory with the standard fluctuation formula results in a capacitance similar to that of the GT model, and the capacitance from the DP-Int dielectric profile is similar to that of GT with $\sigma=5.5$~\AA, like the DP-Bulk model; Fig.~\ref{fig:cex}.
The corrected LMF dielectric profile with Eq.~\ref{eq:lmfeps} or Eq.~\ref{eq:lmfinteq} produces a capacitance in agreement with the LR model, as does the DP-Int model corrected by Eq.~\ref{eq:lmfinteq}.
Similarly, combining the GT model with SPMF or the DP-GT-Bulk MLIP with SPMF both produce capacitance predictions in agreement with that of the LR model, as expected from the inclusion of the longitudinal long range field and its fluctuations in SPMF theory. 

 \section{Conclusions} 
We studied the role of long range electrostatics in determining the nonuniform dielectric tensor of water in the context of empirical and machine learning-based molecular models. 
Short range models reproduce the transverse dielectric profile, $\varepsilon_\parallel(z)$, of the corresponding full model, because long range forces within the plane of the interface effectively cancel and play little role in transverse structure and fluctuations.
In contrast, short range models predict longitudinal dielectric profiles qualitatively different from their long range counterparts. 
We developed a Landau theory to understand this behavior, and show that short range models of water exhibit interfacial correlation lengths approximately an order of magnitude larger than those of the long range models.
The nonuniform dielectric profiles can be related to the bulk dielectric constant, which is the same in both the short and long range models~\cite{Cox2020PNAS,BuiCox2025JPCM,BuiCox2026NatCommun}, but the short range model leads to a bulk value of $\varepsilon_\perp^{-1}(z)$ that is $\varepsilon$ larger than that of the long range model. 
As a result, the capacitance obtained from integrating $\varepsilon_\perp^{-1}(z)$ differs between short and long range models by more than on order of magnitude. 
The structure of short range models can be corrected by learning the averaged forces at the interface, either with LMF theory or through training an MLIP with interfacial configurations~\cite{Niblett2021JCP}. 
In these corrected models, the standard fluctuation formula for $\varepsilon_\perp^{-1}(z)$ must be modified to include the response to the effective field, Eq.~\ref{eq:lmfeps}, to produce dielectric profiles in agreement with long range models. 
We recasted this fluctuation relation as an integral equation, Eq.~\ref{eq:lmfinteq}, that takes as input only polarization fluctuations in the absence of an applied field. 
This integral equation enables accurate predictions of $\varepsilon_\perp^{-1}(z)$ from local short range MLIPs, even those that were not trained on interfacial configurations. 
Alternatively, the linear-scaling SPMF theory~\cite{Hu2014ChemCommun,Yi2017PCCP,Hu2022JCP} adds an instantaneous longitudinal field to the short range systems, including local MLIPs, that leads to the correct estimates of the dielectric constant and capacitance using standard fluctuation formulas. 
Recent developments in classical density function theory (cDFT) combine a learned functional for a short range system with LMF or similar treatments of long range interactions~\cite{Sammuller2023PNAS,BuiCox2024JCP,BuiCox2025PRL,BuiCox2025JPCM,BuiCox2026PNAS}.
Within these frameworks, the integral equation~\ref{eq:lmfinteq} corresponds to using the random phase approximation for the long range part of the direct correlation function, connecting dielectric profiles from cDFT and simulation. 
Building such connections can inform the development of theories for capacitance, ion hydration, electric double layer structure, and hydration interactions~\cite{Bonthuis2011PRL,Bonthuis2012Langmuir,Schneck2012PNAS,Kanduc2014,KanducNetz2015PNAS,Kanduc2016Langmuir,Schlaich2016PRL,Loche2020JPCB}, where a proper description of the induced polarization and dielectric response are critical. 
For example, water polarization-mediated forces between surfaces predicted by short range models would decay on length scales larger than their long range counterparts and differ by orders of magnitude at nanometer separations~\cite{Prelesnik2021PNAS}, exemplifying the need for accurate descriptions of long range interactions in multi-scale theories of dielectric phenomena. 
There are several limitations that we should discuss.
In our theories, we did not take into account the coupling between number density and polarization (or charge density)~\cite{BuiCox2025JPCM,BuiCox2026NatCommun} and does not describe oscillations in the polarization or dielectric functions near the walls.
We also limited our investigation of local MLIPs to a single architecture, DeePMD. 
We expect that similar local MLIPs, like the Behler-Parrinello models~\cite{Ko2021NatCommun,Behler2007PRL} for example, would lead to qualitatively similar results. 
MLIP architectures that inherently capture longer range effects, like graph neural networks, in which message passing enables the description of longer range interactions, are a promising area for extending the analysis performed here. 
In particular, the longer range of graph neural networks should be consistent with a short range model that has a large $\sigma$, on the order of several nanometers or larger. 
However, such models are still inherently short range and unable to quantitatively predict true long wavelength correlations. 
A natural extension of our work is to ionic systems, such as electrolyte solutions, ionic liquids, molten salts, and superionic conductors at interfaces~\cite{Bhargava2008ChemCommun,SalanneMadden2011,Impey1985JCP,Dhattarwal2024NatCommun,Dhattarwal2021Nanoscale,Reed2007JCP,Merlet2013JPCL}
Complete screening of the applied field by mobile ions causes $\varepsilon_\perp^{-1}(z)$ to vanish in the bulk, as required by the Stillinger-Lovett moment conditions~\cite{StillingerLovett1968}.
Short range models of ionic fluids violate perfect screening~\cite{BuiCox2025PRL}, and within the Landau theory developed here, the susceptibility diverges ($\chi_0^{-1}\rightarrow 0$), as does the short range correlation length, $\lambda_0$.
As a result, the standard fluctuation formula should give a parabolic profile for all $L$. 
Static effective fields that account for the effects of long range interactions can correct the structure of ionic fluids and restore perfect screening~\cite{ChenKaurWeeks2004,Rodgers2006PRL,BuiCox2025PRL}, suggesting that our general framework can be extended to ionic systems.
These corrections could be combined with local MLIPs for ionic systems~\cite{Zhang2022NatCommun,Zhang2023PRL,Dhattarwal2025CPC,Dhattarwal2025JCP} to predict double layer structure and capacitance~\cite{Limmer2013,Merlet2014JPCC,Scalfi2020PCCP,Scalfi2021}, for example, with a description of ionic polarization and its significant impact on structure and dynamics~\cite{MaddenWilson1996,Wilson1996JPC,LanningMadden2004,SalanneMadden2011,Dhattarwal2024NatCommun,Dhattarwal2025JCP}.

\appendix
 
 \section{Fluctuation Relation for the Longitudinal Dielectric Constant from LMF Theory}
To determine the nonuniform dielectric constant from LMF theory, we write the dielectric constant as
\begin{equation}
\varepsilon_\perp^{-1}(z) = 1-\frac{\avg{\Delta m(z)}_D}{D}
\end{equation}
and develop a linear response theory for the change in polarization induced by $D$.
The change in polarization, $\Delta m(z)$, is the difference between the polarization of the system in the presence of $D$ and that in the absence of $D$. 
Within LMF theory, the self-consistent field $\Vrl(z)$ also changes, and this change needs to be accounted for in the linear response theory. 
When $D$ is finite, we will refer to the total energy of the renormalized part of the field as
\begin{equation}
\Phi_{\rm R}(\Rbar) = A\int dz \rhoq(z;\Rbar) \Vrl(z;D),
\end{equation}
where we have explicitly indicated the dependence of $\Vrl$ on $D$. 
When $D=0$, the total energy of the renormalized part of the field is
\begin{equation}
\Phi_{\rm R0}(\Rbar) = A\int dz \rhoq(z;\Rbar) \Vrl(z;D=0),
\end{equation}
where the subscript ${\rm R0}$ indicates that the renormalized LMF potential was determined with $D=0$. 
Then, we define
\begin{equation}
\Delta \Phi(\Rbar) \equiv \Phi_{\rm R}(\Rbar) - \Phi_{\rm R0}(\Rbar)
\end{equation}
as the difference between these energies evaluated in the same configuration. 
Our starting point for the linear response theory is the exact expression for the polarization,
\begin{equation}
\avg{m(z;\Rbar)}_{\rm R} = \frac{ \int d\Rbar m(z;\Rbar) e^{-\beta\brac{ U_0(\Rbar) + \Phi_{\rm R}(\Rbar)}} }{ \int d\Rbar e^{-\beta\brac{ U_0(\Rbar) + \Phi_{\rm R}(\Rbar)}} },
\end{equation}
where $U_0(\Rbar)$ is the total energy from short range interactions. 
This expression can be exactly rewritten as an average in the system with no external field through standard techniques~\cite{HuWeeks2010PRL}
\begin{equation}
\avg{m(z;\Rbar)}_{\rm R} = \frac{ \avg{m(z;\Rbar) e^{-\beta\Delta \Phi(\Rbar)}}_{\rm R0} } { \avg{e^{-\beta\Delta \Phi(\Rbar)}}_{\rm R0}}.
\end{equation}
Linearizing this expression yields
\begin{align}
\avg{\Delta m(z)}_D &\equiv  \avg{m(z;\Rbar)}_{\rm R} - \avg{m(z;\Rbar)}_{\rm R0} \\
&\approx  - \beta \avg{\delta m(z;\Rbar) \delta \Delta\Phi(\Rbar)}_{\rm R0},
\label{eq:lmflrt1}
\end{align}
where $\delta m(z;\Rbar)=m(z;\Rbar)-\avg{m(z;\Rbar)}_{\rm R0}$ and $\delta\Delta \Phi(\Rbar) = \Delta \Phi(\Rbar) - \avg{\Delta \Phi(\Rbar)}_{\rm R0}$. 
While this gives us an expression that we can use to evaluate the dielectric constant, we can better connect to the standard fluctuation formula by isolating the contribution from total polarization fluctuations. 
To do so, we note that the total energy contributions from the LMF can be written as
\begin{equation}
\Phi_{\rm R}(\Rbar) = - \frac{D}{\epsilon_0} M(\Rbar) + \frac{A}{\epsilon_0} \int dz \avg{m^\sigma(z;\Rbar)}_{\rm R} m(z;\Rbar)
\end{equation}
and
\begin{equation}
\Phi_{\rm R0}(\Rbar) = \frac{A}{\epsilon_0} \int dz \avg{m^\sigma(z;\Rbar)}_{\rm R0} m(z;\Rbar).
\end{equation}
Inserting these expression for the energies into Eq.~\ref{eq:lmflrt1} gives
\begin{align}
\avg{\Delta m(z)}_D &\approx \beta \frac{D}{\epsilon_0} \avg{\delta m(z;\Rbar) \delta M(\Rbar)}_{\rm R0} \nonumber \\
&- \frac{\beta}{\epsilon_0} \avg{\delta m(z;\Rbar)\brac{\delta \Phi_{\rm S}(\Rbar)-\delta \Phi_{\rm S0}(\Rbar)}}_{\rm R0},
\label{eq:lmfmzd}
\end{align}
where $\Phi_{\rm S}(\Rbar) = A \int dz \avg{m^\sigma(z;\Rbar)}_{\rm R} m(z)$ and $\Phi_{\rm S0}(\Rbar) =A \int dz \avg{m^\sigma(z;\Rbar)}_{\rm R0} m(z)$.
Finally, we can use this linear response expression to write the dielectric constant within LMF theory as the desired fluctuation relation,
\begin{align}
\varepsilon_\perp^{-1}(z) &= 1 - \frac{\beta}{\epsilon_0}  \avg{\delta m(z;\Rbar) \delta M(\Rbar)}_{\rm R0}  \nonumber \\
&+\frac{\beta }{D \epsilon_0} \avg{\delta m(z;\Rbar)\brac{\delta \Phi_{\rm S}(\Rbar)-\delta \Phi_{\rm S0}(\Rbar)}}_{\rm R0}.
\end{align}

\section{Integral Equation for the Longitudinal Dielectric Constant from LMF Theory}
We start with Eq.~\ref{eq:lmfmzd} and write out the LMF potential terms explicitly to obtain
\begin{align}
\big<&\Delta m(z)\big>_D \approx \frac{\beta D A}{\epsilon_0} \int dz' \chi_{\rm R0}(z,z')\nonumber \\
&- \frac{\beta A}{\epsilon_0} \int dz' \chi_{\rm R0}(z,z') \brac{\avg{m^\sigma(z';\Rbar)}_{\rm R} - \avg{m^\sigma(z';\Rbar)}_{\rm R0}} \\
&=\frac{\beta A}{\epsilon_0} \int dz' \chi_{\rm R0}(z,z') \brac{D-\avg{\Delta m^\sigma(z';\Rbar)}_D},
\end{align}
where $\chi_{\rm R0}(z,z')=\avg{\delta m(z;\Rbar)\delta m(z';\Rbar)}_{\rm R0}$.
Then, we can combine $\varepsilon_\perp^{-1}(z) = 1-\avg{\Delta m(z)}_D/D$ with the normalization of the Gaussian distribution to write,
\begin{equation}
D\int dz' \rho_{\rm G}(\len{z-z'})\varepsilon_\perp^{-1}(z')=D-\avg{\Delta m^\sigma(z)}_D.
\end{equation}
which is the term in the square brackets. 
Making this substitution and again using the expression relating $\varepsilon_\perp^{-1}(z)$ and $\avg{\Delta m(z)}_D$ produces the integral equation,
\begin{equation}
\varepsilon_\perp^{-1}(z) = 1- \frac{\beta A}{\epsilon_0} \int dz' \int dz'' \chi_{\rm R0}(z,z') \rho_{\rm G}(\len{z'-z''})\varepsilon_\perp^{-1}(z''),
\end{equation}
which, upon isolating the standard fluctuation formula term, is Eq.~\ref{eq:lmfinteq},
\begin{align}
\varepsilon_\perp^{-1}(z) &= 1 - \frac{\beta}{\epsilon_0} \avg{\delta m(z;\Rbar) \delta M(\Rbar)}_{\rm R0}  \nonumber \\
&+\frac{\beta A}{\epsilon_0} \int dz' \int dz'' \avg{\delta m(z;\Rbar) \delta m(z';\Rbar)}_{\rm R0} \nonumber \\ &\times \rho_G(\len{z'-z''};\sigma)\para{1-\varepsilon_\perp^{-1}(z'')}.
\label{eq:lmfinteq2}
\end{align}
Eq.~\ref{eq:lmfinteq}/\ref{eq:lmfinteq2} is a Fredholm integral equation of the second kind, which can be written in matrix form and solved using standard techniques~\cite{Arfken1985,Press1992}.

\section*{Acknowledgements}
This work was supported by the U.S. Army DEVCOM ARL Army Research Office (ARO) Energy Sciences Competency, Advanced Energy Materials Program award \# W911NF-24-1-0200. The views and conclusions contained in this document are those of the authors and should not be interpreted as representing the official policies, either expressed or implied, of the U.S. Army or the U.S. Government.
We acknowledge the Office of Advanced Research Computing (OARC) at Rutgers,
The State University of New Jersey
for providing access to the Amarel cluster 
and associated research computing resources that have contributed to the results reported here.
ACT acknowledges support from a Zhou Family Fellowship.
We thank Zhonghan Hu (Shandong University) for assistance with the SPMF simulations. 
We thank Steve Cox (Durham University) for insightful discussions and comments on an early draft of the manuscript. 

\bibliography{LRDielectricSlab}

@article{Rodgers2006PRL,
	author = {Rodgers, Jocelyn M. and Kaur, Charanbir and Chen, Yng-Gwei and Weeks, John D.},
	doi = {10.1103/PhysRevLett.97.097801},
	journal = {Phys. Rev. Lett.},
	pages = {097801},
	title = {Attraction Between Like-Charged Walls: Short-Ranged Simulations Using Local Molecular Field Theory},
	volume = {97},
	year = {2006}}

@article{MaddenWilson1996,
	author = {Madden, Paul A. and Wilson, Mark},
	journal = {Chem. Soc. Rev.},
	pages = {339--350},
	title = {`{C}ovalent' effects in `ionic' systems},
	volume = {25},
	year = {1996}}

@article{Reed2007JCP,
	author = {Reed, Stewart K. and Lanning, Oliver J. and Madden, Paul A.},
	journal = {J. Chem. Phys.},
	pages = {084704},
	title = {Electrochemical interface between an ionic liquid and a model metallic electrode},
	volume = {126},
	year = {2007}}

@article{Dhattarwal2025JCP,
	author = {Dhattarwal, Harender S. and Remsing, Richard C.},
	doi = {10.1063/5.0301431},
	journal = {J. Chem. Phys.},
	pages = {234503},
	title = {Electronic fluctuations, cation clustering, and ionic dynamics in molten silver iodide},
	volume = {163},
	year = {2025}}

@article{Dhattarwal2025CPC,
	author = {Dhattarwal, Harender S. and Remsing, Richard C.},
	doi = {10.1002/cphc.202500077},
	journal = {ChemPhysChem},
	pages = {e202500077},
	title = {Electronic Paddlewheels Impact the Dynamics of Superionic Conduction in {AgI}},
	volume = {26},
	year = {2025}}

@article{Dhattarwal2024NatCommun,
	author = {Dhattarwal, Harender S. and Somni, Rahul and Remsing, Richard C.},
	doi = {10.1038/s41467-023-44274-z},
	journal = {Nat. Commun.},
	pages = {121},
	title = {Electronic paddle-wheels in a solid-state electrolyte},
	volume = {15},
	year = {2024}}

@article{StillingerLovett1968,
	author = {Stillinger, Frank H. and Lovett, Ronald},
	doi = {10.1063/1.1670358},
	journal = {J. Chem. Phys.},
	pages = {1991--1994},
	title = {General Restriction on the Distribution of Ions in Electrolytes},
	volume = {49},
	year = {1968}}

@article{Dhattarwal2021Nanoscale,
	author = {Dhattarwal, Harender S. and Remsing, Richard C. and Kashyap, Hemant K.},
	journal = {Nanoscale},
	pages = {4195--4205},
	title = {Intercalation--deintercalation of water-in-salt electrolytes in nanoscale hydrophobic confinement},
	volume = {13},
	year = {2021}}

@article{Bhargava2008ChemCommun,
	author = {Bhargava, B. L. and Balasubramanian, S. and Klein, M. L.},
	doi = {10.1039/B805384G},
	journal = {Chem. Commun.},
	pages = {3339--3351},
	title = {Modelling room temperature ionic liquids},
	year = {2008}}

@article{SalanneMadden2011,
	author = {Salanne, Mathieu and Madden, Paul A.},
	doi = {10.1080/00268976.2011.617523},
	journal = {Mol. Phys.},
	pages = {2299--2315},
	title = {Polarization effects in ionic solids and melts},
	volume = {109},
	year = {2011}}

@article{Wilson1996JPC,
	author = {Wilson, Mark and Madden, Paul A. and Costa-Cabral, Benedito J.},
	doi = {10.1021/jp9512319},
	journal = {J. Phys. Chem.},
	pages = {1227--1237},
	title = {Quadrupole Polarization in Simulations of Ionic Systems: Application to {AgCl}},
	volume = {100},
	year = {1996}}

@article{LanningMadden2004,
	author = {Lanning, Oliver J. and Madden, Paul A.},
	doi = {10.1021/jp048102p},
	journal = {J. Phys. Chem. B},
	pages = {11069--11072},
	title = {Screening at a Charged Surface by a Molten Salt},
	volume = {108},
	year = {2004}}

@article{Merlet2013JPCL,
	author = {Merlet, C{\'e}line and P{\'e}an, Clarisse and Rotenberg, Benjamin and Madden, Paul A. and Simon, Patrice and Salanne, Mathieu},
	doi = {10.1021/jz3019226},
	journal = {J. Phys. Chem. Lett.},
	pages = {264--268},
	title = {Simulating Supercapacitors: Can We Model Electrodes As Constant Charge Surfaces?},
	volume = {4},
	year = {2013}}

@article{Impey1985JCP,
	author = {Impey, R. W. and Klein, M. L. and McDonald, I. R.},
	journal = {J. Chem. Phys.},
	pages = {4690},
	title = {Structural and dynamic properties of lithium sulphate in its solid electrolyte form},
	volume = {82},
	year = {1985}}

@article{Merlet2014JPCC,
	author = {Merlet, C{\'e}line and Limmer, David T. and Salanne, Mathieu and van Roij, Ren{\'e} and Madden, Paul A. and Chandler, David and Rotenberg, Benjamin},
	doi = {10.1021/jp503224w},
	journal = {J. Phys. Chem. C},
	pages = {18291--18298},
	title = {The Electric Double Layer Has a Life of Its Own},
	volume = {118},
	year = {2014}}

@article{Newns1970PRB,
	author = {Newns, D. M.},
	doi = {10.1103/PhysRevB.1.3304},
	journal = {Phys. Rev. B},
	pages = {3304--3322},
	title = {Dielectric Response of a Semi-Infinite Degenerate Electron Gas},
	volume = {1},
	year = {1970}}

@article{Horing1985PRB,
	author = {Horing, Norman J. Morgenstern and Kamen, Elliott and Cui, Hong-Liang},
	doi = {10.1103/PhysRevB.32.2184},
	journal = {Phys. Rev. B},
	pages = {2184--2189},
	title = {Inverse dielectric function of a bounded solid-state plasma},
	volume = {32},
	year = {1985}}

@article{Andersen1972JCP,
	author = {Andersen, Hans C. and Chandler, David},
	doi = {10.1063/1.1678512},
	journal = {J. Chem. Phys.},
	pages = {1918--1929},
	title = {Optimized Cluster Expansions for Classical Fluids. {I}. {G}eneral Theory and Variational Formulation of the Mean Spherical Model and Hard Sphere {P}ercus-{Y}evick Equations},
	volume = {57},
	year = {1972}}

@article{Martin1959PR,
	author = {Martin, Paul C. and Schwinger, Julian},
	doi = {10.1103/PhysRev.115.1342},
	journal = {Phys. Rev.},
	pages = {1342--1373},
	title = {Theory of Many-Particle Systems. {I}},
	volume = {115},
	year = {1959}}

@book{HansenMcDonald2013,
	author = {Hansen, Jean-Pierre and McDonald, Ian R.},
	edition = {4th},
	publisher = {Academic Press},
	title = {Theory of Simple Liquids: With Applications to Soft Matter},
	year = {2013}}

@book{pines2018theory,
	author = {Pines, D.},
	isbn = {9780429972294},
	publisher = {CRC Press},
	title = {Theory Of Quantum Liquids: Normal Fermi Liquids},
	url = {https://books.google.com/books?id=5nhQDwAAQBAJ},
	year = {2018}}

@article{Weeks2002ARPC,
	author = {Weeks, John D.},
	doi = {10.1146/annurev.physchem.53.100201.133929},
	journal = {Annu. Rev. Phys. Chem.},
	pages = {533--562},
	title = {Connecting Local Structure to Interface Formation: A Molecular Scale van der {W}aals Theory of Nonuniform Liquids},
	volume = {53},
	year = {2002}}

@article{Widom1967Science,
	author = {Widom, B.},
	doi = {10.1126/science.157.3787.375},
	journal = {Science},
	number = {3787},
	pages = {375--382},
	title = {Intermolecular Forces and the Nature of the Liquid State},
	volume = {157},
	year = {1967}}

@article{Weeks1971JCP,
	author = {Weeks, John D. and Chandler, David and Andersen, Hans C.},
	doi = {10.1063/1.1674820},
	journal = {J. Chem. Phys.},
	number = {12},
	pages = {5237--5247},
	title = {Role of Repulsive Forces in Determining the Equilibrium Structure of Simple Liquids},
	volume = {54},
	year = {1971}}

@article{Chandler1983Science,
	author = {Chandler, David and Weeks, John D. and Andersen, Hans C.},
	doi = {10.1126/science.220.4599.787},
	journal = {Science},
	number = {4599},
	pages = {787--794},
	title = {Van der {W}aals Picture of Liquids, Solids, and Phase Transformations},
	volume = {220},
	year = {1983}}

@inbook{MaddenChapter,
	author = {Madden, Paul and Kivelson, Daniel},
	booktitle = {Advances in Chemical Physics},
	isbn = {9780470142806},
	pages = {467-566},
	publisher = {John Wiley \& Sons, Ltd},
	title = {A Consistent Molecular Treatment of Dielectric Phenomena},
	year = {1984}}

@article{SprikKlein1988,
	author = {Sprik, Michiel and Klein, Michael L.},
	journal = {J. Chem. Phys.},
	pages = {7556--7560},
	title = {A polarizable model for water using distributed charge sites},
	volume = {89},
	year = {1988}}

@article{Chen2017PNAS,
	author = {Chen, Mohan and Ko, Hsin-Yu and Remsing, Richard C. and Calegari Andrade, Marcos F. and Santra, Biswajit and Sun, Zhaoru and Selloni, Annabella and Car, Roberto and Klein, Michael L. and Perdew, John P. and Wu, Xifan},
	doi = {10.1073/pnas.1712499114},
	journal = {Proc. Natl. Acad. Sci. U.S.A.},
	pages = {10846--10851},
	title = {Ab initio theory and modeling of water},
	volume = {114},
	year = {2017}}

@article{Sun2016NatChem,
	author = {Sun, Jianwei and Remsing, Richard C. and Zhang, Yubo and Sun, Zhaoru and Ruzsinszky, Adrienn and Peng, Haowei and Yang, Zenghui and Paul, Arpita and Waghmare, Umesh and Wu, Xifan and Klein, Michael L. and Perdew, John P.},
	doi = {10.1038/nchem.2535},
	journal = {Nat. Chem.},
	pages = {831--836},
	title = {Accurate first-principles structures and energies of diversely bonded systems from an efficient density functional},
	volume = {8},
	year = {2016}}

@article{HautmanKlein1992,
	author = {Hautman, Joseph and Klein, Michael L.},
	journal = {Mol. Phys.},
	pages = {379--395},
	title = {An {E}wald summation method for planar surfaces and interfaces},
	volume = {75},
	year = {1992}}

@article{Pal2004JCP,
	author = {Pal, Subrata and Balasubramanian, Sundaram and Bagchi, Biman},
	doi = {10.1063/1.1635803},
	journal = {J. Chem. Phys.},
	pages = {1912--1920},
	title = {Anomalous dielectric relaxation of water molecules at the surface of an aqueous micelle},
	volume = {120},
	year = {2004}}

@article{Scalfi2020PCCP,
	author = {Scalfi, Laura and Limmer, David T. and Coretti, Alessandro and Bonella, Sara and Madden, Paul A. and Salanne, Mathieu and Rotenberg, Benjamin},
	doi = {10.1039/C9CP06285H},
	journal = {Phys. Chem. Chem. Phys.},
	pages = {10480--10489},
	title = {Charge fluctuations from molecular simulations in the constant-potential ensemble},
	volume = {22},
	year = {2020}}

@article{Jorgensen1983,
	author = {Jorgensen, William L. and Chandrasekhar, Jayaraman and Madura, Jeffry D. and Impey, Roger W. and Klein, Michael L.},
	doi = {10.1063/1.445869},
	journal = {J. Chem. Phys.},
	pages = {926--935},
	title = {Comparison of simple potential functions for simulating liquid water},
	volume = {79},
	year = {1983}}

@article{Zhang2022NatCommun,
	author = {Zhang, Chunyi and Yue, Shuwen and Panagiotopoulos, Athanassios Z. and Klein, Michael L. and Wu, Xifan},
	journal = {Nat. Commun.},
	pages = {822},
	title = {Dissolving salt is not equivalent to applying a pressure on water},
	volume = {13},
	year = {2022}}

@article{Bhullar2021JMCA,
	author = {Bhullar, Ravneet K. and Zdilla, Michael J. and Klein, Michael L. and Remsing, Richard C.},
	doi = {10.1039/D0TA09635K},
	journal = {J. Mater. Chem. A},
	pages = {6924--6932},
	title = {Effect of water frustration on water oxidation catalysis in the nanoconfined interlayers of layered manganese oxides birnessite and buserite},
	volume = {9},
	year = {2021}}

@article{Watanabe1989,
	author = {Watanabe, Kyoko and Klein, Michael L.},
	journal = {Chem. Phys.},
	pages = {157--167},
	title = {Effective pair potentials and the properties of water},
	volume = {131},
	year = {1989}}

@article{KanducNetz2015PNAS,
	author = {Kandu{\v{c}}, Matej and Netz, Roland R.},
	doi = {10.1073/pnas.1504919112},
	journal = {Proc. Natl. Acad. Sci. U.S.A.},
	pages = {12338--12343},
	title = {From hydration repulsion to dry adhesion between asymmetric hydrophilic and hydrophobic surfaces},
	volume = {112},
	year = {2015}}

@article{Remsing2015JPCL,
	author = {Remsing, Richard C. and McKendry, Ian G. and Strongin, Daniel R. and Klein, Michael L. and Zdilla, Michael J.},
	doi = {10.1021/acs.jpclett.5b02277},
	journal = {J. Phys. Chem. Lett.},
	pages = {4804--4808},
	title = {Frustrated Solvation Structures Can Enhance Electron Transfer Rates},
	volume = {6},
	year = {2015}}

@article{Zhang2021JPCB,
	author = {Zhang, Chunyi and Tang, Fujie and Chen, Mohan and Xu, Jianhang and Zhang, Linfeng and Qiu, Diana Y. and Perdew, John P. and Klein, Michael L. and Wu, Xifan},
	journal = {J. Phys. Chem. B},
	pages = {11444--11456},
	title = {Modeling Liquid Water by Climbing up {J}acob's Ladder in Density Functional Theory Facilitated by Using Deep Neural Network Potentials},
	volume = {125},
	year = {2021}}

@article{Balasubramanian1996,
	author = {Balasubramanian, Sundaram and Mundy, Christopher J. and Klein, Michael L.},
	doi = {10.1063/1.472918},
	journal = {J. Chem. Phys.},
	pages = {11190--11195},
	title = {Shear viscosity of polar fluids: Molecular dynamics calculations of water},
	volume = {105},
	year = {1996}}

@article{Balasubramanian2002JPCB,
	author = {Balasubramanian, Sundaram and Bagchi, Biman},
	doi = {10.1021/jp0133788},
	journal = {J. Phys. Chem. B},
	pages = {3668--3672},
	title = {Slow Orientational Dynamics of Water Molecules at a Micellar Surface},
	volume = {106},
	year = {2002}}

@article{RemsingKlein2017CPL,
	author = {Remsing, Richard C. and Klein, Michael L.},
	journal = {Chem. Phys. Lett.},
	pages = {478},
	title = {Solvation dynamics in water confined within layered manganese dioxide},
	volume = {683},
	year = {2017}}

@article{Kanduc2016Langmuir,
	author = {Kandu{\v{c}}, Matej and Schlaich, Alexander and Schneck, Emanuel and Netz, Roland R.},
	doi = {10.1021/acs.langmuir.6b01727},
	journal = {Langmuir},
	pages = {8767--8782},
	title = {Water-Mediated Interactions between Hydrophilic and Hydrophobic Surfaces},
	volume = {32},
	year = {2016}}

@article{Zhang2023PRL,
	author = {Zhang, Chunyi and Yue, Shuwen and Panagiotopoulos, Athanassios Z. and Klein, Michael L. and Wu, Xifan},
	doi = {10.1103/PhysRevLett.131.076801},
	journal = {Phys. Rev. Lett.},
	pages = {076801},
	title = {Why Dissolving Salt in Water Decreases Its Dielectric Permittivity},
	volume = {131},
	year = {2023}}

@inbook{Arfken1985,
	address = {Orlando, FL},
	author = {Arfken, George},
	edition = {3rd},
	publisher = {Academic Press},
	title = {Mathematical Methods for Physicists},
	year = {1985}}

@inbook{Press1992,
	address = {Cambridge, England},
	author = {Press, William H. and Teukolsky, Saul A. and Vetterling, William T. and Flannery, Brian P.},
	chapter = {18.1},
	edition = {2nd},
	isbn = {0-521-43064-X},
	pages = {782--785},
	publisher = {Cambridge University Press},
	title = {Numerical Recipes in {FORTRAN}: The Art of Scientific Computing},
	year = {1992}}

@article{dhattarwal2023dielectric,
	author = {Dhattarwal, Harender S and Gao, Ang and Remsing, Richard C},
	journal = {The Journal of Physical Chemistry B},
	number = {16},
	pages = {3663--3671},
	publisher = {ACS Publications},
	title = {Dielectric Saturation in Water from a Long-Range Machine Learning Model},
	volume = {127},
	year = {2023}}

@article{Stern2003,
	author = {Stern, Harry A. and Feller, Scott E.},
	doi = {10.1063/1.1537244},
	journal = {J. Chem. Phys.},
	pages = {3401--3412},
	title = {Calculation of the dielectric permittivity profile for a nonuniform system: Application to a lipid bilayer simulation},
	volume = {118},
	year = {2003}}

@article{Ballenegger2005,
	author = {Ballenegger, V. and Hansen, J.-P.},
	doi = {10.1063/1.1845431},
	journal = {J. Chem. Phys.},
	pages = {114711},
	title = {Dielectric permittivity profiles of confined polar fluids},
	volume = {122},
	year = {2005}}

@article{Bonthuis2011PRL,
	author = {Bonthuis, Douwe Jan and Gekle, Stephan and Netz, Roland R.},
	doi = {10.1103/PhysRevLett.107.166102},
	journal = {Phys. Rev. Lett.},
	pages = {166102},
	title = {Dielectric profile of interfacial water and its effect on double-layer capacitance},
	volume = {107},
	year = {2011}}

@article{Bonthuis2012Langmuir,
	author = {Bonthuis, Douwe Jan and Gekle, Stephan and Netz, Roland R.},
	doi = {10.1021/la2051564},
	journal = {Langmuir},
	pages = {7679--7694},
	title = {Profile of the static permittivity tensor of water at interfaces: Consequences for capacitance, hydration interaction and ion adsorption},
	volume = {28},
	year = {2012}}

@article{Schlaich2016PRL,
	author = {Schlaich, Alexander and Knapp, Ernst W. and Netz, Roland R.},
	doi = {10.1103/PhysRevLett.117.048001},
	journal = {Phys. Rev. Lett.},
	pages = {048001},
	title = {Water dielectric effects in planar confinement},
	volume = {117},
	year = {2016}}

@article{Loche2020JPCB,
	author = {Loche, Philip and Ayaz, Cihan and Wolde-Kidan, Amanuel and Schlaich, Alexander and Netz, Roland R.},
	doi = {10.1021/acs.jpcb.0c01967},
	journal = {J. Phys. Chem. B},
	pages = {4365--4371},
	title = {Universal and nonuniversal aspects of electrostatics in aqueous nanoconfinement},
	volume = {124},
	year = {2020}}

@article{Fumagalli2018,
	author = {Fumagalli, L. and Esfandiar, A. and Fabregas, R. and Hu, S. and Ares, P. and Janardanan, A. and Yang, Q. and Radha, B. and Taniguchi, T. and Watanabe, K. and Gomila, G. and Novoselov, K. S. and Geim, A. K.},
	doi = {10.1126/science.aat4191},
	journal = {Science},
	pages = {1339--1342},
	title = {Anomalously low dielectric constant of confined water},
	volume = {360},
	year = {2018}}

@article{Deissenbeck2021,
	author = {Dei{\ss}enbeck, F. and Freysoldt, C. and Todorova, M. and Neugebauer, J. and Wippermann, S.},
	doi = {10.1103/PhysRevLett.126.136803},
	journal = {Phys. Rev. Lett.},
	pages = {136803},
	title = {Dielectric properties of nanoconfined water: A canonical thermopotentiostat approach},
	volume = {126},
	year = {2021}}

@article{Gonella2021,
	author = {Gonella, Grazia and Backus, Ellen H. G. and Nagata, Yuki and Bonthuis, Douwe J. and Loche, Philip and Schlaich, Alexander and Netz, Roland R. and K{\"u}hnle, Angelika and McCrum, Ian T. and Koper, Marc T. M. and Wolf, Martin and Winter, Bernd and Meijer, Gerard and Campen, R. Kramer and Bonn, Mischa},
	doi = {10.1038/s41570-021-00293-2},
	journal = {Nat. Rev. Chem.},
	pages = {466--485},
	title = {Water at charged interfaces},
	volume = {5},
	year = {2021}}

@article{Banuelos2023,
	author = {Ba{\~n}uelos, Jos{\'e} Leobardo and Borguet, Eric and Brown, Gordon E. and Cygan, Randall T. and DeYoreo, James J. and Dove, Patricia M. and Gaigeot, Marie-Pierre and Geiger, Franz M. and Gibbs, Julianne M. and Grassian, Vicki H. and Ilgen, Anastasia G. and Jun, Young-Shin and Kabengi, Nadine and Katz, Lynn and Kubicki, James D. and L{\"u}tzenkirchen, Johannes and Putnis, Christine V. and Remsing, Richard C. and Rosso, Kevin M. and Rother, Gernot and Sulpizi, Marialore and Villalobos, Mario and Zhang, Huichun},
	doi = {10.1021/acs.chemrev.2c00130},
	journal = {Chem. Rev.},
	pages = {6413--6544},
	title = {Oxide- and silicate-water interfaces and their roles in technology and the environment},
	volume = {123},
	year = {2023}}

@article{Limmer2013,
	author = {Limmer, David T. and Merlet, C{\'e}line and Salanne, Mathieu and Chandler, David and Madden, Paul A. and van Roij, Ren{\'e} and Rotenberg, Benjamin},
	doi = {10.1103/PhysRevLett.111.106102},
	journal = {Phys. Rev. Lett.},
	pages = {106102},
	title = {Charge fluctuations in nanoscale capacitors},
	volume = {111},
	year = {2013}}

@article{Scalfi2021,
	author = {Scalfi, Laura and Salanne, Mathieu and Rotenberg, Benjamin},
	doi = {10.1146/annurev-physchem-090519-024042},
	journal = {Annu. Rev. Phys. Chem.},
	pages = {189--212},
	title = {Molecular simulation of electrode-electrolyte interfaces},
	volume = {72},
	year = {2021}}

@article{Marcelja1976,
	author = {Mar{\v{c}}elja, S. and Radi{\'c}, N.},
	journal = {Chem. Phys. Lett.},
	pages = {129--130},
	title = {Repulsion of interfaces due to boundary water},
	volume = {42},
	year = {1976}}

@article{Schneck2012PNAS,
	author = {Schneck, Emanuel and Sedlmeier, Felix and Netz, Roland R.},
	doi = {10.1073/pnas.1205811109},
	journal = {Proc. Natl. Acad. Sci. U.S.A.},
	pages = {14405--14409},
	title = {Hydration repulsion between biomembranes results from an interplay of dehydration and depolarization},
	volume = {109},
	year = {2012}}

@article{Kanduc2014,
	author = {Kandu{\v{c}}, Matej and Schlaich, Alexander and Schneck, Emanuel and Netz, Roland R.},
	doi = {10.1016/j.cis.2014.02.001},
	journal = {Adv. Colloid Interface Sci.},
	pages = {142--152},
	title = {Hydration repulsion between membranes and polar surfaces: Simulation approaches versus continuum theories},
	volume = {208},
	year = {2014}}

@article{Prelesnik2021PNAS,
	author = {Prelesnik, Jesse L. and Alberstein, Robert G. and Zhang, Shuai and Pyles, Harley and Baker, David and Pfaendtner, Jim and De Yoreo, James J. and Tezcan, F. Akif and Remsing, Richard C. and Mundy, Christopher J.},
	doi = {10.1073/pnas.2025121118},
	journal = {Proc. Natl. Acad. Sci. U.S.A.},
	pages = {e2025121118},
	title = {Ion-dependent protein-surface interactions from intrinsic solvent response},
	volume = {118},
	year = {2021}}

@book{Limmer2024Book,
	address = {Oxford},
	author = {Limmer, David T.},
	doi = {10.1093/oso/9780198919858.001.0001},
	publisher = {Oxford University Press},
	title = {Statistical Mechanics and Stochastic Thermodynamics: A Textbook on Modern Approaches in and out of Equilibrium},
	year = {2024}}

@article{Chandler1993PRE,
	author = {Chandler, David},
	doi = {10.1103/PhysRevE.48.2898},
	journal = {Phys. Rev. E},
	pages = {2898--2905},
	title = {Gaussian field model of fluids with an application to polymeric fluids},
	volume = {48},
	year = {1993}}

@article{deLeeuw1980,
	author = {de Leeuw, S. W. and Perram, J. W. and Smith, E. R.},
	doi = {10.1098/rspa.1980.0135},
	journal = {Proc. R. Soc. London A},
	pages = {27--56},
	title = {Simulation of electrostatic systems in periodic boundary conditions. {I}. {L}attice sums and dielectric constants},
	volume = {373},
	year = {1980}}

@article{Cox2020PNAS,
	author = {Cox, Stephen J.},
	doi = {10.1073/pnas.2005847117},
	journal = {Proc. Natl. Acad. Sci. U.S.A.},
	pages = {19746--19752},
	title = {Dielectric response with short-ranged electrostatics},
	volume = {117},
	year = {2020}}

@article{ChenKaurWeeks2004,
	author = {Chen, Yng-Gwei and Kaur, Charanbir and Weeks, John D.},
	doi = {10.1021/jp0469261},
	journal = {J. Phys. Chem. B},
	pages = {19874--19884},
	title = {Connecting systems with short and long ranged interactions: Local molecular field theory for ionic fluids},
	volume = {108},
	year = {2004}}

@article{RodgersWeeks2008JPCM,
	author = {Rodgers, Jocelyn M. and Weeks, John D.},
	doi = {10.1088/0953-8984/20/49/494206},
	journal = {J. Phys.: Condens. Matter},
	pages = {494206},
	title = {Local molecular field theory for the treatment of electrostatics},
	volume = {20},
	year = {2008}}

@article{RodgersWeeks2008PNAS,
	author = {Rodgers, Jocelyn M. and Weeks, John D.},
	doi = {10.1073/pnas.0807623105},
	journal = {Proc. Natl. Acad. Sci. U.S.A.},
	pages = {19136--19141},
	title = {Interplay of local hydrogen-bonding and long-ranged dipolar forces in simulations of confined water},
	volume = {105},
	year = {2008}}

@article{HuWeeks2010PRL,
	author = {Hu, Zhonghan and Weeks, John D.},
	doi = {10.1103/PhysRevLett.105.140602},
	journal = {Phys. Rev. Lett.},
	pages = {140602},
	title = {Efficient solutions of self-consistent mean field equations for dewetting and electrostatics in nonuniform liquids},
	volume = {105},
	year = {2010}}

@article{RodgersHuWeeks2011,
	author = {Rodgers, Jocelyn M. and Hu, Zhonghan and Weeks, John D.},
	journal = {Mol. Phys.},
	pages = {1195--1211},
	title = {On the efficient and accurate short-ranged simulations of uniform polar molecular liquids},
	volume = {109},
	year = {2011}}

@article{RemsingWeeks2015JPCB,
	author = {Remsing, Richard C. and Weeks, John D.},
	journal = {J. Phys. Chem. B},
	pages = {9268--9277},
	title = {Hydrophobicity scaling of aqueous interfaces by an electrostatic mapping},
	volume = {119},
	year = {2015}}

@article{Remsing2016PNAS,
	author = {Remsing, Richard C. and Liu, Shule and Weeks, John D.},
	doi = {10.1073/pnas.1521570113},
	journal = {Proc. Natl. Acad. Sci. U.S.A.},
	pages = {2819--2826},
	title = {Long-ranged contributions to solvation free energies from theory and short-ranged models},
	volume = {113},
	year = {2016}}

@article{Gao2023JPCB,
	author = {Gao, Ang and Remsing, Richard C. and Weeks, John D.},
	journal = {J. Phys. Chem. B},
	pages = {809--821},
	title = {Local molecular field theory for {C}oulomb interactions in aqueous solutions},
	volume = {127},
	year = {2023}}

@article{Hu2014ChemCommun,
	author = {Hu, Zhonghan},
	doi = {10.1039/C4CC06618A},
	journal = {Chem. Commun.},
	pages = {14397--14400},
	title = {Symmetry-preserving mean field theory for electrostatics at interfaces},
	volume = {50},
	year = {2014}}

@article{Yi2017PCCP,
	author = {Yi, Shasha and Pan, Cong and Hu, Liming and Hu, Zhonghan},
	journal = {Phys. Chem. Chem. Phys.},
	pages = {18514--18518},
	title = {On the connections and differences among three mean-field approximations: a stringent test},
	volume = {19},
	year = {2017}}

@article{Hu2022JCP,
	author = {Hu, Zhonghan},
	doi = {10.1063/5.0078007},
	journal = {J. Chem. Phys.},
	pages = {034111},
	title = {The symmetry-preserving mean field condition for electrostatic correlations in bulk},
	volume = {156},
	year = {2022}}

@article{Behler2007PRL,
	author = {Behler, J{\"o}rg and Parrinello, Michele},
	doi = {10.1103/PhysRevLett.98.146401},
	journal = {Phys. Rev. Lett.},
	pages = {146401},
	title = {Generalized neural-network representation of high-dimensional potential-energy surfaces},
	volume = {98},
	year = {2007}}

@article{Bartok2010PRL,
	author = {Bart{\'o}k, Albert P. and Payne, Mike C. and Kondor, Risi and Cs{\'a}nyi, G{\'a}bor},
	doi = {10.1103/PhysRevLett.104.136403},
	journal = {Phys. Rev. Lett.},
	pages = {136403},
	title = {Gaussian approximation potentials: The accuracy of quantum mechanics, without the electrons},
	volume = {104},
	year = {2010}}

@article{Zhang2018PRL,
	author = {Zhang, Linfeng and Han, Jiequn and Wang, Han and Car, Roberto and E, Weinan},
	doi = {10.1103/PhysRevLett.120.143001},
	journal = {Phys. Rev. Lett.},
	pages = {143001},
	title = {Deep potential molecular dynamics: A scalable model with the accuracy of quantum mechanics},
	volume = {120},
	year = {2018}}

@article{Wang2018CPC,
	author = {Wang, Han and Zhang, Linfeng and Han, Jiequn and E, Weinan},
	doi = {10.1016/j.cpc.2018.03.016},
	journal = {Comput. Phys. Commun.},
	pages = {178--184},
	title = {{DeePMD-kit}: A deep learning package for many-body potential energy representation and molecular dynamics},
	volume = {228},
	year = {2018}}

@article{Zeng2023JCP,
	author = {Zeng, Jinzhe and Zhang, Duo and Lu, Denghui and Mo, Pinghui and Li, Zeyu and Chen, Yixiao and Rynik, Mari{\'a}n and Huang, Li'ang and Li, Ziyao and Shi, Shaochen and Wang, Yingze and Ye, Haotian and Tuo, Ping and Yang, Jiabin and Ding, Ye and Li, Yifan and Tisi, Davide and Zeng, Qiyu and Bao, Han and Xia, Yu and Huang, Jiameng and Muraoka, Koki and Wang, Yibo and Chang, Junjie and Yuan, Fengbo and Bore, Sigbj{\o}rn L{\o}land and Cai, Chun and Lin, Yinnian and Wang, Bo and Xu, Jiayan and Zhu, Jia-Xin and Luo, Chenxing and Zhang, Yuzhi and Goodall, Rhys E. A. and Liang, Wenshuo and Singh, Anurag Kumar and Yao, Sikai and Zhang, Jingchao and Wentzcovitch, Renata and Han, Jiequn and Liu, Jie and Jia, Weile and York, Darrin M. and E, Weinan and Car, Roberto and Zhang, Linfeng and Wang, Han},
	doi = {10.1063/5.0155600},
	journal = {J. Chem. Phys.},
	pages = {054801},
	title = {{DeePMD-kit} v2: A software package for deep potential models},
	volume = {159},
	year = {2023}}

@article{Yue2021JCP,
	author = {Yue, Shuwen and Muniz, Maria Carolina and Calegari Andrade, Marcos F. and Zhang, Linfeng and Car, Roberto and Panagiotopoulos, Athanassios Z.},
	doi = {10.1063/5.0031215},
	journal = {J. Chem. Phys.},
	pages = {034111},
	title = {When do short-range atomistic machine-learning models fall short?},
	volume = {154},
	year = {2021}}

@article{Niblett2021JCP,
	author = {Niblett, Samuel P. and Galib, Mirza and Limmer, David T.},
	journal = {J. Chem. Phys.},
	pages = {164101},
	title = {Learning intermolecular forces at liquid-vapor interfaces},
	volume = {155},
	year = {2021}}

@article{Gao2022NatCommun,
	author = {Gao, Ang and Remsing, Richard C.},
	doi = {10.1038/s41467-022-29243-2},
	journal = {Nat. Commun.},
	pages = {1572},
	title = {Self-consistent determination of long-range electrostatics in neural network potentials},
	volume = {13},
	year = {2022}}

@article{Zhang2022DPLR,
	author = {Zhang, Linfeng and Wang, Han and Muniz, Maria Carolina and Panagiotopoulos, Athanassios Z. and Car, Roberto and E, Weinan},
	doi = {10.1063/5.0083669},
	journal = {J. Chem. Phys.},
	pages = {124107},
	title = {A deep potential model with long-range electrostatic interactions},
	volume = {156},
	year = {2022}}

@article{Ko2021NatCommun,
	author = {Ko, Tsz Wai and Finkler, Jonas A. and Goedecker, Stefan and Behler, J{\"o}rg},
	doi = {10.1038/s41467-020-20427-2},
	journal = {Nat. Commun.},
	pages = {398},
	title = {A fourth-generation high-dimensional neural network potential with accurate electrostatics including non-local charge transfer},
	volume = {12},
	year = {2021}}

@article{Grisafi2019JCP,
	author = {Grisafi, Andrea and Ceriotti, Michele},
	doi = {10.1063/1.5128375},
	journal = {J. Chem. Phys.},
	pages = {204105},
	title = {Incorporating long-range physics in local-descriptor-based machine learning: the {LODE} framework},
	volume = {151},
	year = {2019}}

@article{Grisafi2021ChemSci,
	author = {Grisafi, Andrea and Nigam, Jigyasa and Ceriotti, Michele},
	journal = {Chem. Sci.},
	pages = {2078--2090},
	title = {Multi-scale approach for the prediction of atomic scale properties},
	volume = {12},
	year = {2021}}

@article{Anstine2023JPCA,
	author = {Anstine, Dylan M. and Isayev, Olexandr},
	doi = {10.1021/acs.jpca.2c06778},
	journal = {J. Phys. Chem. A},
	pages = {2417--2431},
	title = {Machine learning interatomic potentials and long-range physics},
	volume = {127},
	year = {2023}}

@article{Cheng2025npj,
	author = {Cheng, Bingqing},
	doi = {10.1038/s41524-025-01577-7},
	journal = {npj Comput. Mater.},
	pages = {80},
	title = {Latent {E}wald summation for machine learning of long-range interactions},
	volume = {11},
	year = {2025}}

@article{King2025NatCommun,
	author = {King, Dylan S. and Kim, Dongjin and Zhong, Peichen and Cheng, Bingqing},
	doi = {10.1038/s41467-025-63852-x},
	journal = {Nat. Commun.},
	pages = {8763},
	title = {Machine learning of charges and long-range interactions from energies and forces},
	volume = {16},
	year = {2025}}

@article{Loche2025JCP,
	author = {Loche, Philip and Huguenin-Dumittan, Kevin K. and Honarmand, Melika and Xu, Qianjun and Rumiantsev, Egor and How, Wei Bin and Langer, Marcel F. and Ceriotti, Michele},
	journal = {J. Chem. Phys.},
	pages = {142501},
	title = {Fast and flexible long-range models for atomistic machine learning},
	volume = {162},
	year = {2025}}

@misc{Ramasubramanian2025arXiv,
	archiveprefix = {arXiv},
	author = {Ramasubramanian, Harish and Vazquez-Mayagoitia, Alvaro and Sivaraman, Ganesh and Thakur, Atul C.},
	eprint = {2510.13055},
	note = {arXiv:2510.13055},
	title = {Reciprocal space attention for learning long-range interactions},
	year = {2025}}

@misc{Parker2026arXiv,
	archiveprefix = {arXiv},
	author = {Parker, Isaac J. and Hoffmann, Mandy J. and Baldwin, William J. and Han, Shuang and Gupta, Srishti and Fong, Kara D. and Michaelides, Angelos and Schran, Christoph and De, Sandip and Cs{\'a}nyi, G{\'a}bor},
	eprint = {2603.04228},
	note = {arXiv:2603.04228},
	title = {False metallization in short-ranged machine learned interatomic potentials},
	year = {2026}}

@article{Sammuller2023PNAS,
	author = {Samm{\"u}ller, Florian and Hermann, Sophie and de las Heras, Daniel and Schmidt, Matthias},
	doi = {10.1073/pnas.2312484120},
	journal = {Proc. Natl. Acad. Sci. U.S.A.},
	pages = {e2312484120},
	title = {Neural functional theory for inhomogeneous fluids: Fundamentals and applications},
	volume = {120},
	year = {2023}}

@article{BuiCox2024JCP,
	author = {Bui, Anna T. and Cox, Stephen J.},
	doi = {10.1063/5.0223750},
	journal = {J. Chem. Phys.},
	pages = {104103},
	title = {A classical density functional theory for solvation across length scales},
	volume = {161},
	year = {2024}}

@article{BuiCox2025PRL,
	author = {Bui, Anna T. and Cox, Stephen J.},
	doi = {10.1103/PhysRevLett.134.148001},
	journal = {Phys. Rev. Lett.},
	pages = {148001},
	title = {Learning classical density functionals for ionic fluids},
	volume = {134},
	year = {2025}}

@article{BuiCox2025JPCM,
	author = {Bui, Anna T. and Cox, Stephen J.},
	doi = {10.1088/1361-648X/ade7e7},
	journal = {J. Phys.: Condens. Matter},
	pages = {285101},
	title = {A first-principles approach to electromechanics in liquids},
	volume = {37},
	year = {2025}}

@article{BuiCox2026PNAS,
	author = {Bui, Anna T. and Cox, Stephen J.},
	doi = {10.1073/pnas.2610049123},
	journal = {Proc. Natl. Acad. Sci. U.S.A.},
	pages = {e2610049123},
	title = {A unified machine-learning framework for ab initio multiscale modeling of liquids},
	volume = {123},
	year = {2026}}

@article{BuiCox2026NatCommun,
	author = {Bui, Anna T. and Cox, Stephen J.},
	doi = {10.1038/s41467-026-69482-1},
	journal = {Nat. Commun.},
	pages = {2661},
	title = {Dielectrocapillarity for exquisite control of fluids},
	volume = {17},
	year = {2026}}

@article{Berendsen1987,
	author = {Berendsen, H. J. C. and Grigera, J. R. and Straatsma, T. P.},
	doi = {10.1021/j100308a038},
	journal = {J. Phys. Chem.},
	pages = {6269--6271},
	title = {The missing term in effective pair potentials},
	volume = {91},
	year = {1987}}

@article{Wu2006SPCFw,
	author = {Wu, Yujie and Tepper, Harald L. and Voth, Gregory A.},
	doi = {10.1063/1.2136877},
	journal = {J. Chem. Phys.},
	pages = {024503},
	title = {Flexible simple point-charge water model with improved liquid-state properties},
	volume = {124},
	year = {2006}}

@article{LeeRossky1994,
	author = {Lee, Sang Hyun and Rossky, Peter J.},
	doi = {10.1063/1.466425},
	journal = {J. Chem. Phys.},
	pages = {3334--3345},
	title = {A comparison of the structure and dynamics of liquid water at hydrophobic and hydrophilic surfaces---a molecular dynamics simulation study},
	volume = {100},
	year = {1994}}

@article{Thompson2022LAMMPS,
	author = {Thompson, Aidan P. and Aktulga, H. Metin and Berger, Richard and Bolintineanu, Dan S. and Brown, W. Michael and Crozier, Paul S. and in 't Veld, Pieter J. and Kohlmeyer, Axel and Moore, Stan G. and Nguyen, Trung Dac and Shan, Ray and Stevens, Mark J. and Tranchida, Julien and Trott, Christian and Plimpton, Steven J.},
	doi = {10.1016/j.cpc.2021.108171},
	journal = {Comput. Phys. Commun.},
	pages = {108171},
	title = {{LAMMPS} -- a flexible simulation tool for particle-based materials modeling at the atomic, meso, and continuum scales},
	volume = {271},
	year = {2022}}

@article{Martyna1992,
	author = {Martyna, Glenn J. and Klein, Michael L. and Tuckerman, Mark},
	doi = {10.1063/1.463940},
	journal = {J. Chem. Phys.},
	pages = {2635--2643},
	title = {Nos{\'e}-{H}oover chains: The canonical ensemble via continuous dynamics},
	volume = {97},
	year = {1992}}

@article{YehBerkowitz1999,
	author = {Yeh, In-Chul and Berkowitz, Max L.},
	doi = {10.1063/1.479595},
	journal = {J. Chem. Phys.},
	pages = {3155--3162},
	title = {Ewald summation for systems with slab geometry},
	volume = {111},
	year = {1999}}

@article{ko2025fast,
  title={A fast, accurate, and reactive equivariant foundation potential},
  author={Ko, Tsz Wai and Liu, Runze and Rohan Mishra, Adesh and Yu, Zihan and Qi, Ji and Ong, Shyue Ping},
  journal={arXiv e-prints},
  pages={arXiv--2511},
  year={2025}
}

@article{balantrapu2026neutron,
  title={Neutron and x-ray diffraction reveal the limits of long-range machine learning potentials for medium-range order in silica glass},
  author={Balantrapu, Sai Harshit and Thakur, Atul and Benmore, Chris and Sivaraman, Ganesh},
  journal={Journal of Physics: Materials},
  volume={9},
  number={3},
  pages={035013},
  year={2026},
  publisher={IOP Publishing}
}

\end{document}